\documentclass[lettersize,journal]{IEEEtran}
\usepackage{amsfonts,dsfont,amssymb,amsbsy,amsmath,paralist,theorem,bm,ifthen,color}
\usepackage[pdfstartview=FitH,bookmarksnumbered,unicode,bookmarksopen=true]{hyperref}
\usepackage{graphicx}
\usepackage{epstopdf}
\usepackage{multirow}
\usepackage{amssymb}
\usepackage{subfigure,relsize}
\usepackage{mathrsfs}
\usepackage{bm}
\usepackage{stfloats}
\usepackage{bbm}                
\usepackage{url}
\usepackage[noadjust]{cite}
\usepackage{cases,booktabs}
\usepackage{graphicx,algorithmic,algorithm,relsize}
\usepackage[justification=centering]{caption} 
\usepackage[table]{xcolor}
\usepackage{array}

\newtheorem{rem}{Remark}

\newcommand\psib{\ensuremath{{\bm \psi}}}

\newcommand\Nset  {\ensuremath{{\mathcal{N}}}}
\newcommand\Mset  {\ensuremath{{\mathcal{M}}}}

\newcommand\Hf{\ensuremath{{\mathsf{H}}}}

\graphicspath{{fig/}}

\definecolor{green}{RGB}{34	195	46}
\definecolor{red}{RGB}{220 0 0}

\usepackage{graphicx} 

\title{Generalized Pinching-Antenna Systems: A Radio-Stripe-Based Realization}
\author{Yanqing Xu, \IEEEmembership{Senior Member, IEEE}
         Zhiguo Ding, \IEEEmembership{Fellow, IEEE,}
         and Tsung-Hui Chang, \IEEEmembership{Fellow, IEEE}
         \thanks{\smaller[1] Y. Xu is with the School of Science and Engineering, The Chinese University of Hong Kong, Shenzhen, 518172, China (email: xuyanqing@cuhk.edu.cn).}
         \thanks{\smaller[1] Zhiguo Ding is with the School of Electrical \& Electronic Engineering, Nanyang Technological University, 639798, Singapore  (e-mail: zhiguo.ding@ntu.edu.sg).}
         \thanks{\smaller[1] T.-H. Chang is with the School of Artificial Intelligence, The Chinese University of Hong Kong, Shenzhen, 518172, China (email: changtsunghui@cuhk.edu.cn).} 
         \thanks{\smaller[1] This work has been submitted to the IEEE for possible publication.
         	Copyright may be transferred without notice, after which this version may
         	no longer be accessible.}
}

\begin{document}

\maketitle

\begin{abstract}
    This paper investigates radio stripes (RSs) as a practical realization of generalized pinching antennas and proposes an RS-based generalized pinching-antenna (RS-GPA) framework. Unlike dielectric-waveguide-based passive pinching antennas that rely on passive coupling from a guided wave into free space, RSs employ active antenna processing units (APUs) deployed along a shared cable for local transmission, reception, and signal processing. This cable-like active architecture offers flexible installation and broad frequency applicability, while allowing selected APUs to act as discrete and controllable radiation or reception points for location-flexible wireless access. Based on the proposed RS-GPA framework, we establish the system and channel models by accounting for the distance-dependent APU-user channels. For downlink transmission, we formulate a circuit-power-aware sparse APU activation and beamforming problem and develop a reweighted group-sparse beamforming algorithm. To reveal the activation principle, we analyze the single-user downlink case and characterize when an additional APU should be activated by balancing transmit-power saving and circuit-power cost. Inspired by this insight, a geometry-guided low-complexity multiuser algorithm is proposed. For uplink transmission, we formulate a joint APU activation and user power control problem and develop a geometry-guided sparse activation design. Numerical results show that the proposed RS-GPA framework substantially reduces the total consumed power compared with benchmark schemes, while the geometry-guided algorithm achieves near-identical consumed-power performance to the group-sparse design with significantly lower runtime.
\end{abstract}


\section{Introduction}

Future wireless networks are expected to support diverse services in increasingly complex propagation environments, which calls for flexible-antenna architectures capable of proactively adapting and shaping wireless channel conditions \cite{you2021towards,xu2025distributed,wu2019intelligent,shao20246d,shao20256d,zhu2023modeling,wong2020fluid}. In this context, pinching antennas have recently emerged as a promising flexible-antenna technology, where small dielectric particles are placed on a dielectric waveguide to create reconfigurable radiation points \cite{suzuki2022pinching,ding2025flexible}. Unlike conventional fixed-position antennas, pinching antennas can adjust the locations of radiation points along the waveguide and thus bring wireless access closer to users or target regions. This capability enables effective channel reshaping and propagation-loss reduction, leading to improved spectral efficiency \cite{xu2025rate,qian2026pinching}, enhanced energy efficiency \cite{li2026power,zhou2026spectral}, expanded network coverage \cite{xu2026environment1,xu2026environment2}, and reduced outage probability \cite{tyrovolas2025performance}. Owing to these advantages, pinching antennas have also been investigated in various application scenarios, including integrated sensing and communications (ISAC) \cite{mao2025multi,ouyang2026rate}, non-orthogonal multiple access (NOMA) \cite{xu2025qos,gan2025joint}, energy harvesting \cite{papanikolaou2025resolving,hu2026average}, and physical-layer security \cite{zhong2025physical,zhu2025pinching}.

The concept of pinching antennas has recently been generalized beyond the original dielectric-waveguide implementation \cite{xu2025generalized}. Generalized pinching antennas refer to a broader class of systems in which signals are guided along a physical medium and are selectively radiated or received at configurable locations. This generalized perspective extends the applicability of pinching-antenna principles to diverse deployment scenarios, including indoor and outdoor environments as well as low- and high-frequency communications. Besides dielectric waveguides, other guided-medium structures, such as leaky coaxial cables (LCXs) and metallic pipes, can also create distributed radiation points and thus serve as potential realizations of generalized pinching antennas. In particular, recent studies \cite{wang2025generalized,wang2026leaky1,wang2026leaky2} have shown that LCX-based systems naturally fit this framework, since their radiating slots provide location-dependent access points along the cable. These developments indicate that the essence of pinching antennas is not tied to a specific hardware platform, but rather lies in guided-medium-assisted and location-flexible wireless access. This generalized viewpoint further motivates the exploration of other cable-like distributed antenna infrastructures as potential realizations of generalized pinching-antenna systems.

Radio stripes (RSs) provide a particularly relevant platform for this purpose. An RS is a cable-like distributed antenna architecture in which multiple antenna processing units (APUs) are mounted along a shared cable and connected to a central processing unit (CPU), as depicted in Fig.~1. The shared cable provides data transfer, synchronization, and power supply, while the APUs perform local transmission or reception \cite{interdonato2019ubiquitous}. Since these APUs are spatially distributed along the cable and can be individually controlled, they can naturally serve as discrete and controllable radiation or reception points, making RSs a promising active realization of generalized pinching antennas.

Compared with dielectric-waveguide-based pinching antennas, RS-based generalized pinching antennas (RS-GPAs) offer several practical advantages. First, RSs rely on active APUs rather than passive coupling elements, and thus support local transmission, reception, and signal processing at each access point. Second, cable-connected APUs can be implemented over a broad range of operating frequencies, from sub-6 GHz and millimeter-wave bands to potential terahertz scenarios, depending on the adopted APU hardware. Third, their cable-like form factor, low deployment cost, and flexible installation make RSs suitable for indoor corridors, walls, ceilings, and other structured environments. Owing to these advantages, RSs have been widely studied as practical distributed antenna infrastructures \cite{shaik2021mmse,lopez2022massive,chiotis2024uplink,conceiccao2023access,eberechukwu2025radio}.

Existing studies mainly exploit RSs as distributed antenna and fronthaul infrastructures for scalable cell-free massive MIMO. In these works, the focus is typically on scalable sequential signal processing, fronthaul-efficient implementation, and cooperative transmission or reception among distributed APUs \cite{shaik2021mmse,lopez2022massive,chiotis2024uplink}. In addition, RS selection has been investigated for different objectives. For instance, uplink RS selection was formulated through an RS--user association matrix to jointly optimize spectral efficiency and RS load balancing under sequential maximum ratio combining or optimal sequential linear processing \cite{conceiccao2023access}. RS selection was also studied for user equipment positioning, where a subset of RSs was selected to improve the Cram\'{e}r--Rao lower bound or position error bound under hardware-usage constraints \cite{eberechukwu2025radio}. These studies demonstrate the effectiveness of RSs as practical distributed antenna infrastructures, but they mainly use RSs as low-cost realizations of distributed antenna systems for improving communication efficiency or enhancing localization accuracy.

\begin{figure}[!t]
	\centering
	\includegraphics[width=0.88\linewidth]{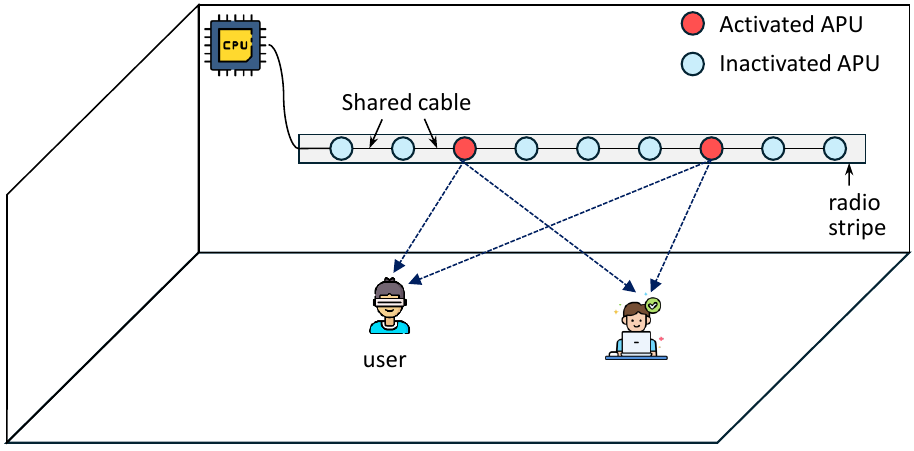}\\
        \captionsetup{justification=justified, singlelinecheck=false, font=small}	
        \caption{Illustration of an RS-GPA system, where selected APUs along a shared cable are activated to create localized radiation or reception points for nearby users.} \label{fig:system model gpa_rs} 
\end{figure}  

Different from these conventional RS studies, this paper relies on the existing RS architecture and exploits its APU-level controllability to realize the key functionality of generalized pinching-antenna systems, namely flexible near-user radiation and reception. From this perspective, activating an APU creates an effective radiation or reception point at a specific location along the stripe, thereby enabling the RS to adapt the wireless access geometry to user locations.
This interpretation leads to a different design perspective. In particular, since the activated APUs may be close to the users and the RS can span a large physical aperture, the APU-user links are characterized by distance-dependent spherical-wave channels. As a result, different APU activation patterns lead to different channel strengths, phase relationships, and multiuser interference structures. Among various possible RS-GPA design objectives, this paper considers energy-efficient downlink and uplink transmission as a representative problem, which captures the fundamental tradeoff between the performance gain of activating more APUs and the corresponding circuit-power cost. The resulting goal is therefore not to activate as many APUs as possible for cooperative processing, but to select a sparse and geometry-aware subset of APUs to form effective near-user radiation or reception points. Motivated by this observation, this paper develops an RS-GPA framework and investigates sparse APU activation for downlink and uplink communications, as depicted in Fig.~\ref{fig:system model gpa_rs}. The main contributions are summarized as follows:
\begin{itemize}
    \item We propose an RS-GPA framework by interpreting the APUs deployed along an RS as discrete and controllable generalized pinching antennas. Different from conventional dielectric-waveguide-based passive pinching antennas, the considered RS-GPA architecture realizes configurable radiation or reception points through active APUs. Based on this architecture, we establish the downlink and uplink signal models by explicitly accounting for the distance-dependent channels between APUs and users.

    \item For downlink transmission, we formulate a circuit-power-aware sparse APU activation and beamforming problem to minimize the total consumed power under user SINR and per-APU power constraints. A reweighted group-sparse beamforming algorithm is developed to jointly determine the active APUs and the beamforming vectors. To better understand the activation principle of RS-GPA systems, we further analyze the single-user downlink case and reveal that an additional APU should be activated only when its transmit-power saving compensates for the corresponding circuit-power cost. Inspired by this insight, we develop a geometry-guided low-complexity multiuser algorithm for sparse APU activation and beamforming.

    \item We further study sparse APU activation for uplink RS-GPA transmission, where active APUs serve as distributed reception points. A joint APU activation and user power control problem is formulated to minimize the total consumed power while satisfying the users' SINR requirements. Based on the uplink signal model, we develop an efficient geometry-aware design that jointly accounts for user transmit power, receive combining gain, distance-dependent channel conditions, and active-APU circuit power.

    \item Numerical results validate the effectiveness of the proposed RS-GPA framework and sparse activation designs. The results show that sparse APU activation substantially reduces the total consumed power compared with full activation, random activation, and fixed antenna baselines. They also demonstrate that the proposed geometry-guided algorithm achieves almost the same consumed-power performance as the group-sparse algorithm with much lower runtime, confirming its suitability for dense RS-GPA deployments.
\end{itemize}

The rest of this paper is organized as follows. Section~\ref{sec: system model} presents the RS-GPA system architecture, including the near-field channel model and the downlink and uplink signal models. Section~\ref{sec: downlink} investigates circuit-power-aware sparse APU activation and beamforming for downlink transmission, including the single-user activation analysis, the reweighted group-sparse beamforming algorithm, and the geometry-guided low-complexity multiuser design. Section~\ref{sec: uplink} studies sparse APU activation and user power control for uplink transmission. Section~\ref{sec: numerical} provides numerical results to validate the effectiveness of the proposed RS-GPA framework and sparse activation algorithms. Finally, Section~\ref{sec: conclusion} concludes the paper.

\section{System Model} \label{sec: system model}

We consider an RS-GPA system, where an RS is deployed to provide location-flexible near-user wireless access, as illustrated in Fig.~\ref{fig:system model gpa_rs}. The RS consists of multiple APUs sequentially mounted along a shared cable and connected to a CPU. Since each APU can be individually controlled and selectively activated, the embedded APUs can be interpreted as discrete generalized pinching antennas. In the downlink, activating an APU is equivalent to creating a controllable radiation point along the stripe.

A three-dimensional Cartesian coordinate system is considered, where a single RS is deployed parallel to the $x$-axis at height $d$, and $N$ APUs are mounted along the stripe. For ease of exposition, we adopt a one-to-one APU--antenna abstraction, where each APU is associated with one antenna element. The position of APU $n$ is denoted by  $\widetilde{\psib}_n = [\tilde{x}_n,0,d]^\top,\quad n\in \Nset \triangleq \{1,\ldots,N\}$, 
where $\tilde{x}_n$ is its horizontal coordinate. Without loss of generality, the APUs are indexed according to their positions along the stripe, i.e., $\tilde{x}_1 < \tilde{x}_2 < \cdots < \tilde{x}_N$.

The system serves $M$ single-antenna users. The position of user $m$ is denoted by $\psib_m = [x_m,y_m,0]^\top,\quad m \in \Mset \triangleq \{1,\ldots,M\}$, 
where $x_m$ and $y_m$ are the horizontal coordinates of user $m$. Accordingly, the distance between APU $n$ and user $m$ is given by
\begin{align}
    r_{m,n}
    = \|\widetilde{\psib}_n-\psib_m\|
    = \sqrt{(\tilde{x}_n-x_m)^2+y_m^2+d^2}.
\end{align}

\subsection{Near-Field Channel Model}

Since the RS can extend over a large physical aperture and the activated APUs may be close to the users, the APU-user links are modeled using a near-field spherical-wave channel. Specifically, the channel from APU $n$ to user $m$ is given by
\begin{equation}
    h_{m,n} = \frac{\sqrt{\beta_0}}{r_{m,n}} \exp\left(-j\frac{2\pi}{\lambda}r_{m,n}\right),
    \label{eq:channel}
\end{equation}
where $\beta_0$ denotes the channel power gain at the reference distance of one meter, $\lambda$ is the free-space carrier wavelength, and $r_{m,n}$ is the exact propagation distance between APU $n$ and user $m$.

Different from the far-field channel models commonly used in conventional RS studies, the channel model in \eqref{eq:channel} explicitly accounts for the location-dependent propagation characteristics of RS-GPA systems. First, the path-loss term $1/r_{m,n}$ reflects the near-user access principle, since activating an APU close to a user can significantly shorten the free-space propagation distance. Second, the phase term $\exp(-j2\pi r_{m,n}/\lambda)$ captures spherical-wave propagation over the large aperture of the RS. As a result, both the amplitude and phase of the user channel depend on the APU location.

\begin{rem} 
    We note that the RS-GPA system differs fundamentally from the dielectric-waveguide-based pinching-antenna system. In particular, in a dielectric-waveguide-based system, multiple pinching antennas on the same waveguide radiate phase-shifted versions of the same information-bearing signal guided along the waveguide \cite{ding2025flexible}. By contrast, each APU in an RS-GPA system is an active transceiver unit connected through the stripe infrastructure and can perform independent signal processing. Therefore, APUs on the same stripe can transmit different data streams and apply different beamforming coefficients. This active processing capability provides additional spatial multiplexing and interference-management degrees of freedom, making the RS-GPA system essentially different from passive waveguide-based pinching-antenna systems.
\end{rem}

\begin{rem}
   It is also worth noting that the channel model of the considered RS-GPA system differs from that of dielectric-waveguide-based pinching-antenna systems due to the different roles of the guided medium in the two architectures. In particular, in dielectric-waveguide systems, a guided radio-frequency (RF) signal propagates along the waveguide and is passively radiated into free space by pinching elements. Hence, the in-waveguide phase shift is part of the end-to-end propagation channel \cite{xu2025pinching}. In contrast, in the considered RS-GPA system, each APU is an active transceiver unit connected to the CPU through the shared cable. The cable mainly provides fronthaul, synchronization, and power supply, rather than serving as a passive RF propagation medium. Therefore, under ideal calibrated fronthaul, the APU-user wireless channel is modeled as the free-space propagation channel between the APU and the user, while cable-induced delays and phase offsets are assumed to be compensated or absorbed into the local precoding or combining coefficients.
\end{rem}

\section{Downlink Transmission Design} \label{sec: downlink}

In this section, we study downlink transmission in the proposed RS-GPA system. The CPU performs centralized precoding and sends the precoded signals to the activated APUs through the shared cable. The activated APUs then radiate the signals to the users, while the inactive APUs remain silent. We first present the downlink signal model and formulate the joint APU activation and beamforming problem. Then, a reweighted group-sparse beamforming algorithm is developed to promote sparse APU activation.

\subsection{Downlink Signal Model and Problem Formulation}

Let $s_m$ denote the information-bearing symbol intended for user $m$, with $\mathbb{E}\{|s_m|^2\}=1$. The signal transmitted by APU $n$ is expressed as
\begin{equation}
    x_n^{\rm DL}
    =
    \sum_{m=1}^{M} w_{n,m}s_m,
\end{equation}
where $w_{n,m}\in\mathbb{C}$ denotes the downlink beamforming coefficient from APU $n$ to user $m$. If APU $n$ is inactive, all its associated beamforming coefficients are forced to zero.

Collecting the beamforming coefficients for user $m$, we define  $\mathbf w_m =
    [w_{1,m},w_{2,m},\ldots,w_{N,m}]^\top \in \mathbb C^{N\times 1}$. 
The overall downlink beamforming matrix is denoted by
\begin{equation}
    \mathbf W =
    [\mathbf w_1,\mathbf w_2,\ldots,\mathbf w_M]\in\mathbb C^{N\times M}.
\end{equation}
Moreover, the beamforming coefficients associated with APU $n$ are collected as
\begin{equation}
    \mathbf v_n =
    [w_{n,1},w_{n,2},\ldots,w_{n,M}]^\top \in \mathbb C^{M\times 1},
\end{equation}
which corresponds to the transpose of the $n$-th row of $\mathbf W$.

Define the channel vector from all APUs to user $m$ as
\begin{equation}
    \mathbf h_m =
    [h_{m,1},h_{m,2},\ldots,h_{m,N}]^\top \in \mathbb C^{N\times 1}.
\end{equation}
The received signal at user $m$ is given by
\begin{align}
    y_m^{\rm DL}
    &=
    \sum_{n=1}^{N} h_{m,n} x_n^{\rm DL} + z_m
    \notag\\
    &=
    \mathbf h_m^\top \mathbf w_m s_m
    +
    \sum_{\ell\neq m}\mathbf h_m^\top \mathbf w_\ell s_\ell
    + z_m,
\end{align}
where $z_m\sim\mathcal{CN}(0,\sigma_m^2)$ denotes the additive white Gaussian noise at user $m$. Accordingly, the downlink signal-to-interference-plus-noise ratio (SINR) of user $m$ is expressed as
\begin{equation}
    \gamma_m^{\rm DL}
    =
    \frac{|\mathbf h_m^\top \mathbf w_m|^2}
    {\sum_{\ell\neq m}|\mathbf h_m^\top \mathbf w_\ell|^2+\sigma_m^2}.
    \label{eq:dl_sinr}
\end{equation}

Under the generalized pinching-antenna setting, an RS can be flexibly deployed near the target service region to support near-field communications, where the distances between APUs and users have a significant impact on the channel strengths. Since many APUs can be mounted along one stripe, different APUs may contribute very differently to the users. Activating all APUs is therefore not always energy efficient, as weakly contributing APUs still consume circuit power and signal-processing resources. This motivates sparse APU activation, where only a subset of APUs are selected for transmission.

Motivated by this, we jointly optimize APU activation and downlink beamforming to minimize the total network power consumption, including both transmit power and the circuit power of activated APUs, while satisfying the users' quality-of-service (QoS) requirements. Let $a_n$ denote the activation variable of APU $n$, where $a_n=1$ if APU $n$ is activated for downlink transmission and $a_n=0$ otherwise. The downlink design problem is formulated as

\begin{subequations}\label{prob:dl_original}
    \begin{align}
        \mathop{\min}_{\{a_n\},\mathbf W}\quad
        &
        \sum_{n=1}^{N}\sum_{m=1}^{M}|w_{n,m}|^2
        +
        P_{\rm c}\sum_{n=1}^{N}a_n
        \label{prob:dl_original_obj}
        \\
        \mathrm{s.t.}\quad
        &
        \gamma_m^{\rm DL}\geq \Gamma_m,\quad m\in\Mset,
        \label{prob:dl_original_sinr}
        \\
        &
        \sum_{m=1}^{M}|w_{n,m}|^2
        \leq a_n P_n^{\max},\quad n\in\Nset,
        \label{prob:dl_original_power}
        \\
        &
        \sum_{n=1}^{N}a_n \geq M,
        \label{prob:dl_original_cardinality}
        \\
        &
        a_n\in\{0,1\},\quad n\in\Nset.
        \label{prob:dl_original_binary}
    \end{align}
\end{subequations}
Here, $P_{\rm c}$ denotes the circuit power consumed by each activated APU, $\Gamma_m$ is the target SINR of user $m$, and $P_n^{\max}$ is the maximum transmit power of APU $n$. Constraint~\eqref{prob:dl_original_power} imposes a per-APU transmit-power limit and couples the activation variables with the beamforming coefficients. Specifically, if $a_n=0$, then $\sum_{m=1}^{M}|w_{n,m}|^2=0$, and APU $n$ is inactive. If $a_n=1$, the total transmit power of APU $n$ is limited by $P_n^{\max}$. Constraint~\eqref{prob:dl_original_cardinality} enforces the minimum number of active APUs required by the adopted multiuser transmission scheme.


It is worth noting that problem \eqref{prob:dl_original} is different from conventional antenna selection in co-located multiple-input multiple-output (MIMO) systems \cite{gao2017massive,asaad2018massive}. In conventional co-located MIMO, each selected antenna is an element of a compact array. In the considered RS-GPA system, however, each active APU corresponds to a distributed access point at a specific location along the RS. Therefore, the activation variable $a_n$ not only determines whether APU $n$ consumes circuit power, but also determines whether the wireless link associated with its particular APU-user distances participates in downlink transmission. Under the spherical-wave channel model in \eqref{eq:channel}, different APU activation patterns lead to different channel strengths, phase relationships, and multiuser interference structures. 
This near-field geometry-dependent feature also distinguishes problem \eqref{prob:dl_original} from conventional RS designs, since APU activation here directly determines which distance-dependent propagation links are included in the downlink transmission design.
Based on this observation, the following subsections reformulate APU activation through the row sparsity of the beamforming matrix and further exploit the single-user activation insight to develop a low-complexity geometry-guided design that prioritizes APUs with favorable user--APU distances.

\subsection{Group-Sparse Beamforming Reformulation}

The key observation is that the activation state of APU $n$ can be characterized by the sparsity of the beamforming row vector $\mathbf v_n$. Specifically, APU $n$ participates in downlink transmission if and only if
\begin{equation}
    \|\mathbf v_n\|_2^2
    =
    \sum_{m=1}^{M}|w_{n,m}|^2
    >0.
\end{equation}
Therefore, the APU activation problem can be interpreted as a group-sparse beamforming problem, where each APU corresponds to one group. Based on this observation, the activation cost can be represented by an $\ell_0$-type row-sparsity penalty, and problem~\eqref{prob:dl_original} can be reformulated as
\begin{subequations} \label{prob:dl_l0}
\begin{align}
    \mathop{\min}_{\mathbf W}\quad
    & \sum_{n=1}^{N}\|\mathbf v_n\|_2^2
    + P_{\rm c}\sum_{n=1}^{N}
    \mathbbm{1}\left(\|\mathbf v_n\|_2^2>0\right)
    \label{prob:dl_l0_obj} \\
    \mathrm{s.t.}\quad
    & \gamma_m^{\rm DL}\geq \Gamma_m,\quad m\in\Mset,
    \label{prob:dl_l0_sinr} \\
    & \|\mathbf v_n\|_2^2 \leq P_n^{\max},\quad n\in\Nset,
    \label{prob:dl_l0_power} \\
    & \sum_{n=1}^{N}
    \mathbbm{1}\left(\|\mathbf v_n\|_2^2>0\right)
    \geq M .
    \label{prob:dl_l0_cardinality}
\end{align}
\end{subequations}
where $\mathbbm{1}(\cdot)$ denotes the indicator function. This reformulation removes the explicit binary variables, but remains non-convex due to the discontinuous $\ell_0$-type sparsity penalty and the lower-bound group-sparsity constraint in~\eqref{prob:dl_l0_cardinality}.

To obtain a tractable approximation, we adopt the reweighted sparsity-promoting technique used in sparse beamforming \cite{dai2014sparse,candes2008enhancing}. Specifically, the activation indicator
$\mathbbm{1}\left(\|\mathbf v_n\|_2^2>0\right)$ is interpreted as an $\ell_0$-type penalty on the group power $\|\mathbf v_n\|_2^2$. Following the reweighted $\ell_1$ approximation, the indicator term is replaced at iteration $t$ by a weighted group-power penalty, i.e.,
\begin{equation}
    \mathbbm{1}\left(\|\mathbf v_n\|_2^2>0\right)
    \approx
    \alpha_n^{(t)}\|\mathbf v_n\|_2^2,
\end{equation}
where
\begin{equation}
    \alpha_n^{(t)} = \frac{1}{\|\mathbf v_n^{(t)}\|_2^2+\epsilon},
    \label{eqn: alpha update}
\end{equation}
and $\epsilon>0$ is a small regularization parameter. This weight update assigns a larger penalty to APUs with smaller transmit power in the previous iteration, thereby encouraging weakly used APUs to be switched off.

With the above reweighted approximation, the original mixed-integer problem can be approximated at iteration $t$ as the following weighted transmit-power minimization problem:
\begin{subequations} \label{prob:dl_weighted}
\begin{align}
    \mathop{\min}_{\mathbf W}\quad
    & \sum_{n=1}^{N}
    \omega_n^{(t)}\|\mathbf v_n\|_2^2
    \label{prob:dl_weighted_obj} \\
    \mathrm{s.t.}\quad
    & \gamma_m^{\rm DL}\geq \Gamma_m,\quad m\in\Mset,
    \label{prob:dl_weighted_sinr} \\
    & \|\mathbf v_n\|_2^2 \leq P_n^{\max},\quad n\in\Nset.
    \label{prob:dl_weighted_power}
\end{align}
\end{subequations}
where the weight coefficient in the objective is given by
\begin{equation}
    \omega_n^{(t)} = 1+P_{\rm c}\alpha_n^{(t)}
    = 1+ \frac{P_{\rm c}}{\|\mathbf v_n^{(t)}\|_2^2+\epsilon}.
    \label{eq:dl_weight}
\end{equation}
It can be observed from \eqref{eq:dl_weight} that an APU with a small transmit power in the previous iteration is assigned a large weight in the current iteration. As a result, the corresponding beamforming row vector is further penalized, which promotes row sparsity of the beamforming matrix and encourages unnecessary APUs to be deactivated.

We note that the minimum active-APU requirement in~\eqref{prob:dl_l0_cardinality} is not directly imposed in the reweighted SOCP subproblem~\eqref{prob:dl_weighted}, since a convex continuous approximation of this lower-bound sparsity constraint is not straightforward. Instead, this requirement is enforced during the active-set extraction and refinement stages by retaining at least $M$ APUs. If the resulting active set is infeasible, inactive APUs are added back according to their row powers until the SINR constraints are satisfied.

For fixed $\{\alpha_n^{(t)},\omega_n^{(t)}\}$, problem \eqref{prob:dl_weighted} is a weighted power minimization problem with SINR constraints and per-APU power constraints. Although the SINR constraints in \eqref{prob:dl_weighted_sinr} are non-convex in their original form, they can be equivalently transformed into second-order cone constraints. Specifically, without loss of optimality, the phase of the desired signal component $\mathbf h_m^\top \mathbf w_m$ can be rotated such that
\begin{equation}
    \mathrm{Im}\{\mathbf h_m^\top\mathbf w_m\}=0,\quad
    \mathrm{Re}\{\mathbf h_m^\top\mathbf w_m\}\geq 0.
\end{equation}
Then, the SINR constraint $\gamma_m^{\rm DL}\geq \Gamma_m$ can be equivalently rewritten as
\begin{equation}
    \left\|
    \begin{bmatrix}
    \mathbf h_m^\top\mathbf w_1 \\
    \cdots \\
    \mathbf h_m^\top\mathbf w_{m-1} \\
    \mathbf h_m^\top\mathbf w_{m+1} \\
    \cdots \\
    \mathbf h_m^\top\mathbf w_M \\
    \sigma_m
    \end{bmatrix}
    \right\|_2
    \leq
    \frac{1}{\sqrt{\Gamma_m}}
    \mathrm{Re}\{\mathbf h_m^\top\mathbf w_m\},
    \quad m\in\Mset.
    \label{eq:dl_soc_sinr}
\end{equation}
Moreover, the per-APU power constraint can be written as
\begin{equation}
    \|\mathbf v_n\|_2 \leq \sqrt{P_n^{\max}},
    \quad n\in\Nset,
    \label{eq:dl_soc_power}
\end{equation}
which is also a second-order cone constraint. Therefore, for fixed weights $\{\omega_n^{(t)}\}$, problem \eqref{prob:dl_weighted} can be equivalently solved as
\begin{equation}
\begin{aligned}
    \mathop{\min}_{\mathbf W}\quad
    & \sum_{n=1}^{N}\omega_n^{(t)}\|\mathbf v_n\|_2^2 \\
    \mathrm{s.t.}\quad
    & \eqref{eq:dl_soc_sinr},\ \eqref{eq:dl_soc_power},\\
    & \mathrm{Im}\{\mathbf h_m^\top\mathbf w_m\}=0,\quad m\in\Mset .
\end{aligned}
\label{prob:dl_soc}
\end{equation}
Problem \eqref{prob:dl_soc} is SOCP-representable and can be efficiently solved using standard convex optimization solvers.

After the reweighted iterations converge, the active APU set is obtained according to the row powers of the optimized beamforming matrix. Specifically, for a small threshold $\delta>0$, we first define
\begin{equation}
    \widetilde{\mathcal A}
    =
    \left\{
    n\in\Nset:
    \|\mathbf v_n\|_2^2>\delta
    \right\}.
    \label{eq:dl_active_set_initial}
\end{equation}
To satisfy the spatial-multiplexing requirement in \eqref{prob:dl_original_cardinality}, the active set is then constructed as
\begin{equation}
    \mathcal A =
    \begin{cases}
    \widetilde{\mathcal A}, & |\widetilde{\mathcal A}|\ge M,\\
    \widetilde{\mathcal A}\cup \mathcal B, & |\widetilde{\mathcal A}|<M,
    \end{cases}
    \label{eq:dl_active_set}
\end{equation}
where $\mathcal B$ contains the $M-|\widetilde{\mathcal A}|$ APUs in $\Nset\setminus\widetilde{\mathcal A}$ with the largest row powers. Thus, the extracted active set always satisfies $|\mathcal A|\ge M$.

With the active set $\mathcal A$ fixed, a final beamforming refinement is performed by solving
\begin{subequations} \label{prob:dl_refine}
\begin{align}
    \mathop{\min}_{\mathbf W}\quad
    & \sum_{n\in\mathcal A}\|\mathbf v_n\|_2^2
    \label{prob:dl_refine_obj} \\
    \mathrm{s.t.}\quad
    & \gamma_m^{\rm DL}\geq \Gamma_m,\quad m\in\Mset,
    \label{prob:dl_refine_sinr} \\
    & \|\mathbf v_n\|_2^2\leq P_n^{\max},\quad n\in\mathcal A,
    \label{prob:dl_refine_power} \\
    & \mathbf v_n=\mathbf 0,\quad n\notin\mathcal A.
    \label{prob:dl_refine_inactive}
\end{align}
\end{subequations}
Problem \eqref{prob:dl_refine} is also an SOCP. If \eqref{prob:dl_refine} is feasible, it yields the final downlink beamforming solution for the selected active APUs. Otherwise, an active-set recovery step is performed by adding back inactive APUs in descending order of their row powers obtained from the reweighted solution, and \eqref{prob:dl_refine} is solved again. This process is repeated until feasibility is recovered. The details are summarized in Algorithm~\ref{alg:dl_group_sparse}.

\begin{algorithm}[t]
\caption{Reweighted Group-Sparse Downlink Beamforming}
\label{alg:dl_group_sparse}\small
\begin{algorithmic}[1]
\STATE Initialize $\mathbf W^{(0)}$ by solving the full-APU transmit-power minimization problem with all APUs available. Set $t=0$.
\REPEAT
    \STATE Compute the row vectors $\mathbf v_n^{(t)}$, $\forall n\in\Nset$.
    \STATE Update $\alpha_n^{(t)},\ n\in\Nset$ according to \eqref{eqn: alpha update}.
    \STATE Update the objective weights $\omega_n^{(t)},\ n\in\Nset$ according to \eqref{eq:dl_weight}.
    \STATE Solve the SOCP form of \eqref{prob:dl_weighted} with fixed $\{\omega_n^{(t)}\}$ and obtain $\mathbf W^{(t+1)}$.
    \STATE Set $t\leftarrow t+1$.
\UNTIL{the objective value converges or the maximum number of iterations is reached}
\STATE Determine the preliminary active set $\widetilde{\mathcal A}$ according to \eqref{eq:dl_active_set_initial}.
\IF{$|\widetilde{\mathcal A}|<M$}
    \STATE Add the $M-|\widetilde{\mathcal A}|$ APUs with the largest row powers in $\Nset\setminus\widetilde{\mathcal A}$ to form $\mathcal A$.
\ELSE
    \STATE Set $\mathcal A=\widetilde{\mathcal A}$.
\ENDIF
\STATE Sort the inactive APUs in $\Nset\setminus\mathcal A$ in descending order of their row powers $\|\mathbf v_n\|_2^2$.
\REPEAT
    \STATE Solve the beamforming refinement problem in \eqref{prob:dl_refine} over the current active set $\mathcal A$.
    \IF{\eqref{prob:dl_refine} is infeasible}
        \STATE Add the inactive APU with the largest row power to $\mathcal A$.
    \ENDIF
\UNTIL{\eqref{prob:dl_refine} is feasible or $\mathcal A=\Nset$}
\STATE Output the active APU set $\mathcal A$ and the beamforming matrix $\mathbf W$.
\end{algorithmic}
\end{algorithm}

The above reweighted group-sparse beamforming algorithm provides a general approach for joint APU activation and downlink beamforming. Its main advantage is that it handles the multiuser SINR constraints, per-APU transmit-power constraints, and activation cost within a unified convex-optimization-based framework. Moreover, since APU activation is induced by the row sparsity of the beamforming matrix, the algorithm avoids explicit binary optimization over the APU activation variables. The minimum active-APU requirement induced by simultaneous multiuser transmission is enforced during active-set construction and refinement, ensuring that the final active set satisfies $|\mathcal A|\ge M$.

However, Algorithm~\ref{alg:dl_group_sparse} still requires solving an SOCP in each reweighted iteration, whose dimension grows with the number of APUs and users. Therefore, when the RS is long and contains a large number of APUs, directly running Algorithm~\ref{alg:dl_group_sparse} may incur high computational complexity. In addition, the generic sparse beamforming formulation does not explicitly exploit the ordered one-dimensional geometry of the RS or the near-field structure of the APU-user channels. To address these issues, we next study the single-user case with $M=1$ to obtain useful activation insights. Based on these insights, a low-complexity geometry-guided multiuser design is then developed.

\subsection{Single-User Energy-Efficient APU Activation Analysis} \label{sec: single user}

Before developing a low-complexity multiuser design, we first consider the single-user case, i.e., $M=1$, to obtain useful insights into APU activation. For notational simplicity, the user index is omitted in this subsection. For a given active APU set $\mathcal A\subseteq\Nset$, the received signal is given by
\begin{equation}
    y = \sum_{n\in\mathcal A} h_n w_n s + z,
\end{equation}
where $w_n$ is the beamforming coefficient of APU $n$, $s$ is the information-bearing symbol with $\mathbb E\{|s|^2\}=1$, and $z\sim\mathcal{CN}(0,\sigma^2)$ is the additive Gaussian noise.

For a fixed active set $\mathcal A$, the minimum transmit power required to satisfy the target SNR $\Gamma$ is achieved by maximum ratio transmission, yielding
\begin{equation}
    P_{\mathcal A}^{\rm req}
    =
    \frac{\Gamma\sigma^2}{\sum_{n\in\mathcal A}|h_n|^2}.
    \label{eq:single_user_req_power_general}
\end{equation}
Under the LoS near-field channel model in \eqref{eq:channel}, we have $|h_n|^2=\beta_0/r_n^2$, where $r_n$ is the distance between APU $n$ and the user. Therefore, we have
\begin{equation}
    P_{\mathcal A}^{\rm req}
    =
    \frac{\Gamma\sigma^2}
    {\beta_0\sum_{n\in\mathcal A} r_n^{-2}}.
    \label{eq:single_user_req_power}
\end{equation}
Taking into account the circuit power consumed by active APUs, the total power consumption is given by
\begin{equation}
    P_{\mathcal A}^{\rm tot}
    =
    \frac{\Gamma\sigma^2}
    {\beta_0\sum_{n\in\mathcal A} r_n^{-2}}
    +
    |\mathcal A|P_{\rm c}.
    \label{eq:single_user_total_power}
\end{equation}

Equation~\eqref{eq:single_user_total_power} reveals two useful observations. First, for a fixed number of active APUs, minimizing \eqref{eq:single_user_total_power} is equivalent to maximizing $\sum_{n\in\mathcal A}r_n^{-2}$. Hence, the optimal active set consists of the APUs with the largest channel gains, or equivalently, the APUs closest to the user. Second, although activating more APUs increases $\sum_{n\in\mathcal A}r_n^{-2}$ and thus reduces the required transmit power, it also incurs additional circuit power consumption. Therefore, the optimal number of active APUs is determined by the tradeoff between transmit-power reduction and APU activation power consumption.

To characterize this tradeoff, let
\begin{equation}
    r_{(1)}\leq r_{(2)}\leq \cdots \leq r_{(N)}
\end{equation}
denote the ordered APU-user distances, and define
\begin{equation}
    G_L
    =
    \sum_{\ell=1}^{L}\frac{1}{r_{(\ell)}^2}.
    \label{eq:GL_definition}
\end{equation}
Then, the minimum total power consumption achieved by activating the $L$ closest APUs is given by
\begin{equation}
    P_L^{\rm tot}
    =
    \frac{\Gamma\sigma^2}{\beta_0 G_L}
    +
    LP_{\rm c},
    \quad L=1,\ldots,N.
    \label{eq:single_user_total_power_L}
\end{equation}
Accordingly, the optimal number of active APUs can be obtained through the following one-dimensional search:
\begin{equation}
    L^\star
    =
    \arg\min_{1\leq L\leq N}
    \left(
    \frac{\Gamma\sigma^2}{\beta_0 G_L}
    +
    LP_{\rm c}
    \right).
    \label{eq:single_user_optimal_L}
\end{equation}

We now characterize when it is beneficial to activate one more APU. Suppose that the $L$ closest APUs have been activated. Activating the $(L+1)$-th closest APU is energy-efficient if and only if
\begin{equation}
    P_{L+1}^{\rm tot}<P_L^{\rm tot}.
\end{equation}
Using \eqref{eq:single_user_total_power_L}, this condition becomes
\begin{equation}
    \frac{\Gamma\sigma^2}{\beta_0 G_{L+1}}
    +
    (L+1)P_{\rm c}
    <
    \frac{\Gamma\sigma^2}{\beta_0 G_L}
    +
    LP_{\rm c}.
\end{equation}
Since $G_{L+1}=G_L+1/r_{(L+1)}^2$, the above inequality can be equivalently expressed as
\begin{equation}
    P_{\rm c}
    <
    \frac{\Gamma\sigma^2}{\beta_0}
    \left(
    \frac{1}{G_L}
    -
    \frac{1}{G_{L+1}}
    \right),
    \label{eq:activation_condition_1}
\end{equation}
or, equivalently,
\begin{equation}
    P_{\rm c}
    <
    \frac{\Gamma\sigma^2}{\beta_0}
    \frac{r_{(L+1)}^{-2}}
    {G_L\left(G_L+r_{(L+1)}^{-2}\right)}.
    \label{eq:activation_condition_2}
\end{equation}

The condition in \eqref{eq:activation_condition_2} provides a useful additional-APU activation rule. The right-hand side represents the transmit-power reduction brought by activating the $(L+1)$-th closest APU. Hence, this APU should be activated only when the resulting transmit-power saving exceeds its circuit-power cost $P_{\rm c}$. If the additional APU is far from the user, $r_{(L+1)}^{-2}$ is small and the corresponding power saving is limited, making its activation inefficient. In contrast, if the additional APU is still close enough to provide a sufficiently large channel-gain increment, its activation can reduce the total power consumption.

This result shows that the optimal number of active APUs is determined by the tradeoff between channel-gain improvement and circuit-power consumption. Therefore, full activation is not necessarily optimal, even though activating more APUs always increases the effective channel gain. Moreover, the active APU selection is directly linked to the user's location through the ordered APU-user distances. For a fixed number of active APUs, the APUs closest to the user are preferred, while condition \eqref{eq:activation_condition_2} determines whether an additional, farther APU is worth activating from an energy-efficiency perspective. This observation is particularly relevant for RS-GPA systems, where a large number of APUs may be deployed along the stripe and sparse activation serves as a discrete form of location-dependent antenna placement for energy-efficient transmission.

\subsection{Low-Complexity Geometry-Guided Multiuser Design}

Motivated by the single-user analysis in subsection \ref{sec: single user}, we further develop a low-complexity geometry-guided sparse APU activation and beamforming algorithm. The analysis shows that APUs closer to the user should be prioritized, while activating an additional APU is beneficial only when the resulting transmit-power reduction can compensate for its circuit-power cost. Extending this idea to the multiuser case, we evaluate each APU according to its distances to all users and first identify a small set of geometry-favorable APUs. The subsequent APU selection and beamforming design are then carried out mainly within this smaller candidate set. In this way, the proposed algorithm reduces the computational complexity while still following the distance-dependent activation principle revealed by the single-user analysis.

Specifically, for APU $n$, we define the geometry score as
\begin{equation}
    q_n
    =
    \sum_{m=1}^{M}
    \frac{1}{r_{m,n}^2+\epsilon_{\rm g}},
    \label{eq:geometry_score}
\end{equation}
where $r_{m,n}=\|\boldsymbol \psi_m-\widetilde{\boldsymbol \psi}_n\|$ denotes the distance between user $m$ and APU $n$, and $\epsilon_{\rm g}>0$ is a small constant introduced to avoid numerical instability. The score in \eqref{eq:geometry_score} is inspired by the distance-dependent power-saving behavior revealed in the previous subsection. A larger $q_n$ indicates that APU $n$ is, in general, closer to the users and is therefore more likely to provide useful channel gain for downlink transmission. Based on this score, the APUs are sorted in descending order of $\{q_n\}$, and a geometry-guided candidate pool $\mathcal C$ is formed by selecting the first $C_{\rm pool}$ APUs, where $C_{\rm pool}$ is a design parameter satisfying
\begin{equation}
    M \leq C_{\rm pool} \leq N .
\end{equation}
By choosing $C_{\rm pool}$ to be much smaller than $N$, the subsequent active-set search and beamforming optimization can be performed over a reduced set of APUs, thereby lowering the computational complexity.

After constructing the candidate pool, the active APU set is initialized in a user-centric manner. For each user, its nearest APU within $\mathcal C$ is first selected. If the resulting active set contains fewer than $M$ APUs, additional APUs are selected from $\mathcal C$ according to the geometry-priority order until at least $M$ APUs are activated. This initialization follows the simultaneous multiuser beamforming requirement and provides sufficient spatial degrees of freedom for serving multiple users.

Given an active set $\mathcal A$, the beamforming vectors are obtained by solving the following fixed-active-set power minimization problem:
\begin{subequations}\label{prob:dl_fixed_active_alg2}
\begin{align}
    \mathop{\min}_{\mathbf W}\quad
    & \sum_{n\in\mathcal A}\|\mathbf v_n\|_2^2
    \label{prob:dl_fixed_active_alg2_obj} \\
    \mathrm{s.t.}\quad
    & \gamma_m^{\rm DL}\geq \Gamma_m,\quad m\in\Mset,
    \label{prob:dl_fixed_active_alg2_sinr} \\
    & \|\mathbf v_n\|_2^2\leq P_n^{\max},\quad n\in\mathcal A,
    \label{prob:dl_fixed_active_alg2_power} \\
    & \mathbf v_n=\mathbf 0,\quad n\notin\mathcal A .
    \label{prob:dl_fixed_active_alg2_inactive}
\end{align}
\end{subequations}
Problem~\eqref{prob:dl_fixed_active_alg2} can be transformed into an SOCP using the same SINR reformulation as in \eqref{eq:dl_soc_sinr}. If the initial active set is infeasible, APUs in $\mathcal C\setminus\mathcal A$ are sequentially added according to the geometry-priority order until a feasible solution is obtained. If the candidate pool is insufficient to achieve feasibility, a conservative fallback step is used by adding APUs outside $\mathcal C$ according to the global geometry-priority order. This step improves the robustness of the algorithm, while the main search process remains geometry-guided and low-complexity when the candidate pool is sufficient.

Once a feasible active set is obtained, a restricted forward addition step is performed within the candidate pool. Specifically, inactive APUs in $\mathcal C\setminus\mathcal A$ are tested one by one according to the geometry-priority order. An APU is added only if the resulting fixed-active-set beamforming solution reduces the total consumed power
\begin{equation}
    P_{\rm total}
    =
    \sum_{n\in\mathcal A}\|\mathbf v_n\|_2^2
    +
    P_{\rm c}|\mathcal A|.
\end{equation}
Finally, a one-pass backward pruning step is applied, where weakly used APUs are tested for removal according to their beamforming row powers. An APU is removed only if the resulting active set still satisfies $|\mathcal A|\ge M$, the SINR constraints remain feasible, and the total consumed power decreases. Therefore, the proposed method preserves the key insight from the single-user analysis, namely that activating an additional APU is beneficial only when the associated transmit-power saving can compensate for its circuit-power cost.

The proposed geometry-guided algorithm is summarized in Algorithm~\ref{alg:geometry_guided_dl}.

\begin{algorithm}[t]
\caption{Low-Complexity Geometry-Guided Sparse APU Activation and Beamforming}
\label{alg:geometry_guided_dl}\small
\begin{algorithmic}[1]
\STATE Compute the geometry score $q_n$ for each APU according to \eqref{eq:geometry_score}.
\STATE Sort the APUs in descending order of $\{q_n\}$ and construct the candidate pool $\mathcal C$ with size $C_{\rm pool}$.
\STATE Initialize the active set $\mathcal A$ by selecting, for each user, its nearest APU in $\mathcal C$.
\IF{$|\mathcal A|<M$}
    \STATE Add APUs from $\mathcal C$ according to the geometry-priority order until $|\mathcal A|\geq M$.
\ENDIF
\STATE Solve the fixed-active-set beamforming problem in \eqref{prob:dl_fixed_active_alg2}.
\WHILE{\eqref{prob:dl_fixed_active_alg2} is infeasible and $\mathcal A\neq\mathcal C$}
    \STATE Add the highest-priority APU in $\mathcal C\setminus\mathcal A$ to $\mathcal A$.
    \STATE Re-solve \eqref{prob:dl_fixed_active_alg2}.
\ENDWHILE
\IF{\eqref{prob:dl_fixed_active_alg2} is still infeasible}
    \STATE Add APUs outside $\mathcal C$ according to the global geometry-priority order until feasibility is achieved or all APUs are included.
\ENDIF
\STATE Perform restricted forward addition within $\mathcal C$: add an inactive APU only if it reduces the total consumed power.
\STATE Perform one-pass backward pruning: remove a weakly used APU only if $|\mathcal A|\geq M$ after removal, the SINR constraints remain feasible, and the total consumed power decreases.
\STATE Output the active APU set $\mathcal A$ and the beamforming matrix $\mathbf W$.
\end{algorithmic}
\end{algorithm}

Compared with Algorithm~\ref{alg:dl_group_sparse}, Algorithm~\ref{alg:geometry_guided_dl} avoids solving reweighted sparse beamforming problems over all APUs. Instead, it relies on a geometry-guided candidate pool and solves fixed-active-set beamforming problems for a small number of candidate active sets. When the candidate pool is sufficient, its computational burden is mainly governed by the candidate-pool size $C_{\rm pool}$ rather than the total number of deployed APUs $N$. Meanwhile, since the beamforming vectors are still optimized under the multiuser SINR constraints, the proposed algorithm retains the ability to account for multiuser interference after the geometry-based candidate screening.

\section{Uplink Transmission Design} \label{sec: uplink}

In this section, we study the uplink transmission design for the proposed RS-GPA system. Different from the downlink case, where the activated APUs jointly transmit signals to the users, in the uplink the activated APUs receive the users' signals and forward the corresponding baseband observations to the CPU for centralized detection. We focus on the centralized uplink design and assume that the fronthaul capacity of the RS is sufficiently large to collect the received signals from the activated APUs at the CPU. Fronthaul-aware uplink processing, such as quantized observation forwarding and fronthaul-constrained APU activation, is an important practical extension and will be investigated in future work. The design objective is to jointly optimize APU activation, receive combining, and user transmit powers to minimize the total power consumption while satisfying the users' SINR requirements.

\subsection{Uplink Signal Model and Problem Formulation}

Let $p_m$ denote the transmit power of user $m$, and let $s_m$ denote its transmitted symbol with $\mathbb E\{|s_m|^2\}=1$. For a given active APU set $\mathcal A\subseteq\Nset$, the received signal at the activated APUs is given by
\begin{equation}
    \mathbf y_{\mathcal A}^{\rm UL}
    =
    \sum_{m=1}^{M}\sqrt{p_m}\mathbf h_m(\mathcal A)s_m
    +
    \mathbf z_{\mathcal A},
    \label{eq:ul_received_signal}
\end{equation}
where $\mathbf h_m(\mathcal A)=[h_{m,n}]_{n\in\mathcal A}$ denotes the channel vector from user $m$ to the activated APUs, and $\mathbf z_{\mathcal A}\sim\mathcal{CN}(\mathbf 0,\sigma^2\mathbf I)$ denotes the receiver noise vector at the activated APUs.

The CPU applies a receive combining vector $\mathbf u_m\in\mathbb C^{|\mathcal A|\times 1}$ to detect user $m$'s signal, i.e.,
\begin{equation}
    \hat{s}_m
    =
    \mathbf u_m^\Hf\mathbf y_{\mathcal A}^{\rm UL}.
\end{equation}
Accordingly, the uplink SINR of user $m$ is given by
\begin{equation}
    \gamma_m^{\rm UL}
    =
    \frac{
    p_m|\mathbf u_m^\Hf\mathbf h_m(\mathcal A)|^2
    }
    {
    \sum_{\ell\neq m}p_\ell|\mathbf u_m^\Hf\mathbf h_\ell(\mathcal A)|^2
    +
    \sigma^2\|\mathbf u_m\|_2^2
    }.
    \label{eq:ul_sinr}
\end{equation}

The uplink power consumption consists of the user transmit power and the circuit power consumed by the activated APUs. The joint uplink APU activation, power control, and receive combining problem is formulated as
\begin{subequations}\label{prob:ul_original}
\begin{align}
    \mathop{\min}_{\mathcal A,\{p_m\},\{\mathbf u_m\}}\quad
    &
    \sum_{m=1}^{M}p_m
    +
    |\mathcal A|P_{\rm c}
    \label{prob:ul_original_obj}\\
    \mathrm{s.t.}\quad
    &
    \gamma_m^{\rm UL}\geq \Gamma_m,\quad m\in\Mset,
    \label{prob:ul_original_sinr}\\
    &
    0\leq p_m\leq P_m^{\max},\quad m\in\Mset,
    \label{prob:ul_original_power}\\
    &
    |\mathcal A|\geq M,
    \label{prob:ul_original_min_apu}\\
    &
    \mathcal A\subseteq\Nset.
    \label{prob:ul_original_set}
\end{align}
\end{subequations}
Constraint~\eqref{prob:ul_original_min_apu} enforces the minimum number of active APUs required by the adopted simultaneous multiuser uplink reception scheme.

Different from the downlink case, we do not formulate uplink APU activation through the row sparsity of a receive combining matrix. This is because receive combiners have arbitrary scaling, and their norms are not direct physical indicators of APU activation. Instead, we explicitly optimize the active APU set. For a given active set, the receive combining and user power control can be handled efficiently, as described next.

\subsection{Uplink Power Control for a Given Active APU Set}

For a fixed active APU set $\mathcal A$ satisfying $|\mathcal A|\geq M$, the remaining design variables are the receive combiners $\{\mathbf u_m\}$ and the user transmit powers $\{p_m\}$. For given transmit powers, the receive combining vector of each user can be updated using the MMSE receiver. Specifically, for fixed $\{p_m\}$, the MMSE combining vector for user $m$ is given by
\begin{equation}
    \mathbf u_m
    =
    \left(
    \sum_{\ell=1}^{M}p_\ell\mathbf h_\ell(\mathcal A)\mathbf h_\ell^\Hf(\mathcal A)
    +
    \sigma^2\mathbf I
    \right)^{-1}
    \mathbf h_m(\mathcal A).
    \label{eq:ul_mmse_combiner}
\end{equation}
The scaling of $\mathbf u_m$ does not affect the SINR in \eqref{eq:ul_sinr}; hence, the above expression can be used without normalization.

For fixed receive combiners, the uplink SINR constraints can be transformed into power-control constraints. Specifically, $\gamma_m^{\rm UL}\geq\Gamma_m$ is equivalent to
\begin{equation}
    p_m
    \geq
    \sum_{\ell\neq m}
    \frac{\Gamma_m|\mathbf u_m^\Hf\mathbf h_\ell(\mathcal A)|^2}
    {|\mathbf u_m^\Hf\mathbf h_m(\mathcal A)|^2}
    p_\ell
    +
    \frac{\Gamma_m\sigma^2\|\mathbf u_m\|_2^2}
    {|\mathbf u_m^\Hf\mathbf h_m(\mathcal A)|^2},
    \quad m\in\Mset.
    \label{eq:ul_power_constraint}
\end{equation}
Define the nonnegative matrix $\mathbf F(\mathcal A)\in\mathbb R_+^{M\times M}$ and vector $\boldsymbol\eta(\mathcal A)\in\mathbb R_+^{M\times 1}$ as
\begin{equation}
    [\mathbf F(\mathcal A)]_{m,\ell}
    =
    \begin{cases}
    \dfrac{\Gamma_m|\mathbf u_m^\Hf\mathbf h_\ell(\mathcal A)|^2}
    {|\mathbf u_m^\Hf\mathbf h_m(\mathcal A)|^2}, & \ell\neq m,\\
    0, & \ell=m,
    \end{cases}
    \label{eq:ul_F}
\end{equation}
and
\begin{equation}
    \eta_m(\mathcal A)
    =
    \frac{\Gamma_m\sigma^2\|\mathbf u_m\|_2^2}
    {|\mathbf u_m^\Hf\mathbf h_m(\mathcal A)|^2}.
    \label{eq:ul_eta}
\end{equation}
Then the uplink power-control constraints can be compactly written as
\begin{equation}
    \mathbf p\geq \mathbf F(\mathcal A)\mathbf p+\boldsymbol\eta(\mathcal A),
    \label{eq:ul_power_vector}
\end{equation}
where $\mathbf p=[p_1,\ldots,p_M]^\top$. For fixed receive combiners, if $\rho(\mathbf F(\mathcal A))<1$, where $\rho(\cdot)$ denotes the spectral radius, the minimum power vector satisfying \eqref{eq:ul_power_vector} is
\begin{equation}
    \mathbf p^\star(\mathcal A)
    =
    \left(\mathbf I-\mathbf F(\mathcal A)\right)^{-1}\boldsymbol\eta(\mathcal A).
    \label{eq:ul_power_solution}
\end{equation}
If $\rho(\mathbf F(\mathcal A))\geq1$ or some entry of $\mathbf p^\star(\mathcal A)$ exceeds the corresponding maximum transmit power, the active set $\mathcal A$ is declared infeasible for the given receive combiners.

Since the MMSE combiners and the user transmit powers are coupled, we alternately update the receive combiners using \eqref{eq:ul_mmse_combiner} and the power vector using \eqref{eq:ul_power_solution} until convergence or until infeasibility is detected. The resulting total uplink power consumption for active set $\mathcal A$ is denoted by
\begin{equation}
    P_{\rm UL}(\mathcal A)
    =
    \sum_{m=1}^{M}p_m^\star(\mathcal A)
    +
    |\mathcal A|P_{\rm c}.
    \label{eq:ul_total_power_set}
\end{equation}
This fixed-active-set uplink procedure serves as a building block for the APU activation design.

\subsection{Low-Complexity Geometry-Guided Uplink Design}
Solving \eqref{prob:ul_original} by exhaustive search over all possible active APU sets is computationally prohibitive when $N$ is large. To reduce the search space, we develop a low-complexity geometry-guided uplink design. The motivation can be understood from the single-user uplink case. For one uplink user and a given active APU set $\mathcal A$, maximum-ratio combining yields
\begin{equation}
    \gamma^{\rm UL}
    =
    \frac{
    p\sum_{n\in\mathcal A}|h_n|^2
    }{\sigma^2},
\end{equation}
where $p$ is the user transmit power. Therefore, the minimum transmit power required to satisfy the target SNR $\Gamma$ is
\begin{equation}
    p_{\mathcal A}^{\rm req}
    =
    \frac{\Gamma\sigma^2}
    {\sum_{n\in\mathcal A}|h_n|^2}
    =
    \frac{\Gamma\sigma^2}
    {\beta_0\sum_{n\in\mathcal A}r_n^{-2}},
\end{equation}
where the second equality follows from the near-field channel model in \eqref{eq:channel}. This expression has the same distance-dependent structure as the single-user downlink result. Hence, activating APUs closer to the user can reduce the required user transmit power, whereas activating additional APUs also incurs circuit-power consumption. This provides an uplink counterpart to the downlink activation principle.

Motivated by this observation, instead of searching over all APUs, we first identify a small set of geometry-favorable APUs and then perform uplink power control and receive combining over this reduced candidate pool. Specifically, APUs far away from the users usually provide limited useful channel gain, and thus their activation may bring only marginal user-transmit-power reduction while still incurring circuit-power consumption.

For APU $n$, we define the uplink geometry score as
\begin{equation}
    q_n^{\rm UL}
    =
    \sum_{m=1}^{M}
    \frac{1}{r_{m,n}^2+\epsilon_{\rm g}},
    \label{eq:ul_geometry_score}
\end{equation}
where $r_{m,n}=\|\boldsymbol\psi_m-\widetilde{\boldsymbol\psi}_n\|$ denotes the distance between user $m$ and APU $n$, and $\epsilon_{\rm g}>0$ is a small constant introduced to avoid numerical instability. A larger value of $q_n^{\rm UL}$ indicates that APU $n$ is geometrically closer to the users and is therefore more likely to provide useful receive signal power. This score is consistent with the distance-dependent channel gain of the RS-GPA channel and the marginal activation principle discussed above.

We sort all APUs in descending order of $\{q_n^{\rm UL}\}$ and construct a candidate pool $\mathcal C$ containing the first $C_{\rm pool}^{\rm UL}$ APUs, where
\begin{equation}
    M\leq C_{\rm pool}^{\rm UL}\leq N.
\end{equation}
The candidate-pool size $C_{\rm pool}^{\rm UL}$ is a design parameter that controls the tradeoff between performance and complexity. A larger candidate pool provides more spatial diversity for uplink reception, while a smaller candidate pool reduces the number of active-set trials.

After constructing $\mathcal C$, the active set is initialized in a user-centric manner. Specifically, for each user, its nearest APU within $\mathcal C$ is selected. If the resulting active set contains fewer than $M$ APUs, additional APUs are added from $\mathcal C$ according to the geometry-priority order until $|\mathcal A|\geq M$. The fixed-active-set uplink procedure described in the previous subsection is then applied to this active set. If the active set is infeasible, APUs in $\mathcal C\setminus\mathcal A$ are sequentially added according to the geometry-priority order until feasibility is achieved. If the candidate pool is insufficient to obtain a feasible solution, a conservative fallback step is applied by adding APUs outside $\mathcal C$ according to the global geometry-priority order.

Once a feasible active set is obtained, a restricted forward addition step is performed within $\mathcal C$. Specifically, inactive APUs in $\mathcal C\setminus\mathcal A$ are tested one by one according to the geometry-priority order. An APU is added only if the resulting fixed-active-set uplink design reduces the total uplink power consumption in \eqref{eq:ul_total_power_set}. Finally, a one-pass backward pruning step is performed, where APUs with small receive-combining contribution are tested for removal. An APU is removed only if the resulting active set still satisfies $|\mathcal A|\geq M$, remains feasible, and yields a lower total uplink power consumption.

Compared with exhaustive active-set search, this geometry-guided method substantially reduces the search space by exploiting the geometry-induced channel structure of the RS. Compared with a purely nearest-APU rule, it still accounts for multiuser interference through the fixed-active-set MMSE combining and power-control updates. Therefore, the proposed geometry-guided uplink design provides a practical low-complexity approach for joint APU activation, receive combining, and user power control.

\section{Numerical Results} \label{sec: numerical}

In this section, numerical simulations are conducted to evaluate the performance of the proposed RS-GPA system and the effectiveness of the proposed algorithms. To highlight the complementary role of RS-GPA systems, we consider a representative sub-$6$ GHz scenario where the carrier frequency is set to $f_{\rm c}=3.5$ GHz. This is different from dielectric-waveguide-based pinching-antenna systems, which have been mainly applied for high-frequency scenarios. Unless otherwise specified, $N=12$ APUs are uniformly deployed along an RS of length $60$ m, and the rectangular communication area is set to $60$ m $\times$ $10$ m, and the noise power is set to $\sigma_m^2=-70$ dBm, $\forall m$. The number of users is set to $M=3$, and the target SINR is set to $\Gamma_m=0$ dB for all users unless otherwise specified. For the proposed reweighted group-sparse beamforming algorithm, the iteration terminates when the relative change of the objective value between two consecutive iterations is below $10^{-4}$. All simulation results are obtained by averaging over $50$ independent random realizations of user positions.

\subsection{Performance Evaluation for Downlink Transmission}

\begin{figure}[!t]
	\centering
	\includegraphics[width=0.92\linewidth]{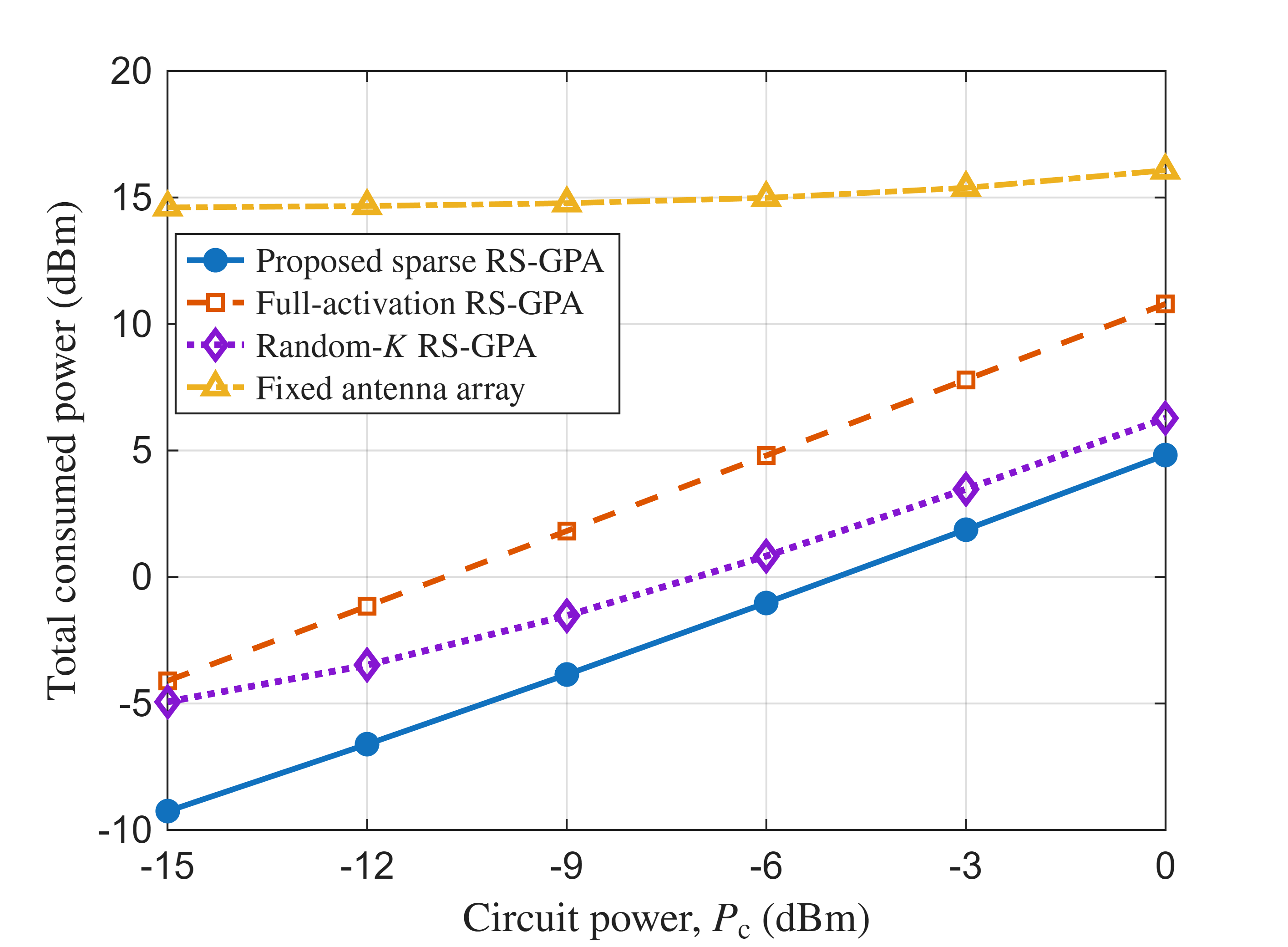}\\
        \captionsetup{justification=justified, singlelinecheck=false, font=small}	
        \caption{Total consumed power versus the circuit power $P_{\rm c}$ for different downlink transmission schemes.} \label{fig:pc_power}  
\end{figure}

We first evaluate the proposed RS-GPA architecture and the corresponding downlink design algorithms. For the proposed sparse RS-GPA scheme, the reweighted group-sparse beamforming method in Algorithm~\ref{alg:dl_group_sparse} is adopted to jointly determine the active APUs and the downlink beamforming vectors. Three benchmark schemes are considered for comparison: 1) \emph{Full-activation RS-GPA}, where all APUs on the RS are activated and the beamforming vectors are optimized; 2) \emph{Random-$K$ RS-GPA}, where $K$ APUs are randomly activated and the beamforming vectors are optimized, with $K$ set equal to the number of APUs selected by the proposed scheme; and 3) \emph{Fixed antenna array}, where $M$ conventional antennas are placed at the center of the service region with half-wavelength spacing and the beamforming vectors are optimized. For all schemes, the total consumed power includes both the transmit power and the circuit power of the active transmitting units.
Fig.~\ref{fig:pc_power} shows the total consumed power versus the circuit power $P_{\rm c}$ in the considered deployment. It can be observed that the consumed power of all schemes increases with $P_{\rm c}$, since activating each transmitting unit becomes more costly. The full-activation RS-GPA scheme benefits from the cooperative beamforming gain provided by all distributed APUs, but its consumed power increases rapidly as $P_{\rm c}$ grows because all APUs remain active and incur circuit-power consumption. The fixed antenna array consumes the highest power over the whole considered range, since its fixed antenna locations cannot exploit the spatial flexibility of the RS and may be far from many users.
In contrast, the proposed sparse RS-GPA scheme consistently achieves the lowest consumed power. Compared with the fixed antenna array, the proposed scheme achieves a substantial power reduction over the entire considered range, with more than $20$ dB gain observed in the low-$P_{\rm c}$ regime and more than $10$ dB gain maintained even when $P_{\rm c}$ is high. This confirms that sparse APU activation can effectively select favorable APU locations and reduce the required transmit power. Compared with the full-activation RS-GPA scheme, the proposed scheme avoids unnecessary APU activation when circuit power is non-negligible. Compared with the random-$K$ RS-GPA scheme, the proposed scheme still achieves clear power savings even when the same number of APUs is activated, which verifies the importance of optimized APU selection.

\begin{figure}[!t]
	\centering
	\includegraphics[width=0.92\linewidth]{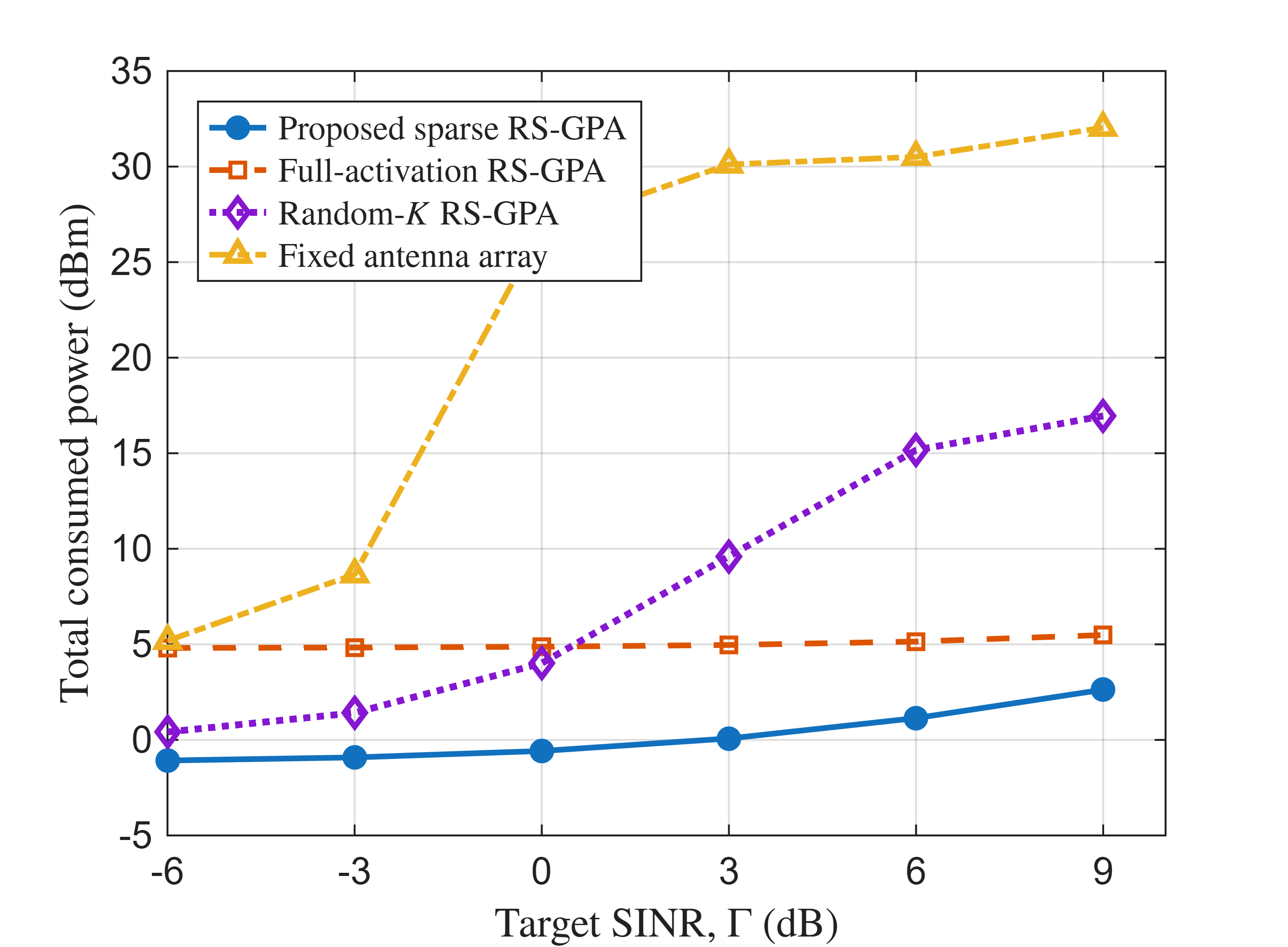}\\
        \captionsetup{justification=justified, singlelinecheck=false, font=small}	
        \caption{Total consumed power versus target SINR $\Gamma$ for different downlink transmission schemes.} \label{fig:power_vs_sinr} 
\end{figure}

Fig.~\ref{fig:power_vs_sinr} shows the total consumed power versus the target SINR $\Gamma$. As expected, the consumed power generally increases with $\Gamma$, since a more stringent QoS requirement requires higher transmit power and/or more active APUs. The proposed sparse RS-GPA scheme achieves the lowest consumed power over the entire considered SINR range. This is because it jointly selects energy-efficient APUs and optimizes the beamforming vectors, thereby avoiding unnecessary circuit-power consumption while satisfying the SINR constraints.
The random-$K$ RS-GPA scheme performs relatively close to the proposed scheme when the SINR target is low, but its consumed power increases rapidly as $\Gamma$ becomes larger. This indicates that, under stringent SINR requirements, merely activating the same number of APUs is insufficient; the locations of the activated APUs become crucial for reducing transmit power and managing multiuser interference. The full-activation RS-GPA scheme exhibits an almost steady consumed power because all APUs are always activated and the total consumed power is dominated by the circuit-power consumption. However, it still consumes more power than the proposed sparse design, since many APUs are unnecessarily activated. The fixed antenna array suffers from the highest consumed power, because its fixed antenna locations cannot exploit the spatial flexibility provided by the RS. These results confirm the effectiveness of sparse APU activation and beamforming optimization for energy-efficient downlink RS-GPA transmission.

\begin{figure*}[!t]
	\centering
	\includegraphics[width=0.98\linewidth]{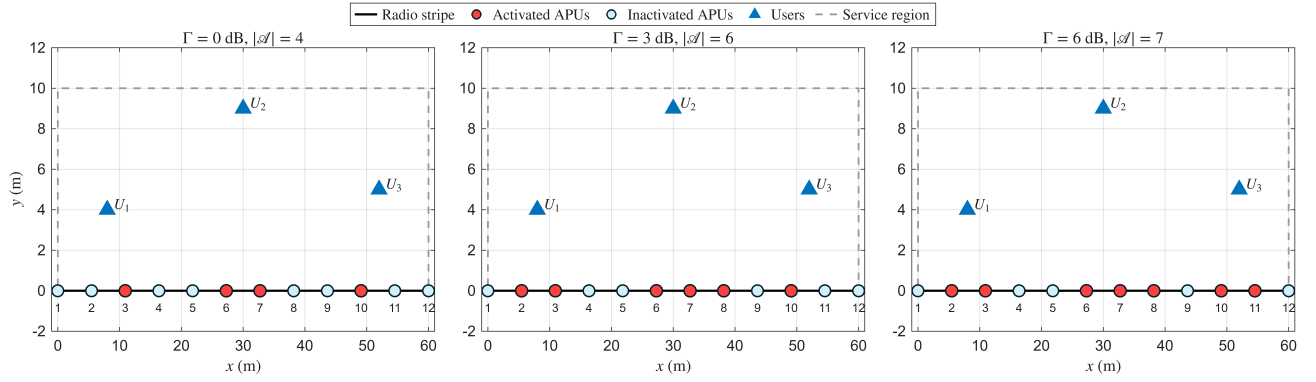}\\
        \captionsetup{justification=justified, singlelinecheck=false, font=small}	
        \caption{Top-view visualization of the active APU selection under different SINR requirements, where the users' locations are fixed, $P_c = -12$ dBm and the target SINR is set to $\Gamma=0$, $3$, and $6$ dB, respectively.}  \label{fig:top_view_active_apus}
\end{figure*} 

Fig.~\ref{fig:top_view_active_apus} provides a top-view visualization of the active APU selection results under different SINR requirements. The users' locations are fixed, while the target SINR is set to $\Gamma=0$, $3$, and $6$ dB, respectively. It can be observed that the number of active APUs increases as the target SINR becomes more stringent. Specifically, $4$ APUs are activated when $\Gamma=0$ dB, while $6$ and $7$ APUs are activated when $\Gamma=3$ dB and $\Gamma=6$ dB, respectively. This is because a higher SINR requirement requires stronger received signal power and more effective multiuser interference management, which can be supported by activating additional APUs along the RS.
It is also observed that the activated APUs are mainly selected around the users' projected locations on the RS. This behavior is consistent with the distance-dependent channel characteristics of RS-GPA systems, where APUs closer to the users generally provide more useful channel gains. Therefore, Fig.~\ref{fig:top_view_active_apus} intuitively demonstrates the adaptive activation capability of the proposed scheme: fewer APUs are sufficient for low QoS requirements, while more APUs are activated when stronger beamforming capability is needed.

\begin{figure}[!t]
	\centering
	\includegraphics[width=0.92\linewidth]{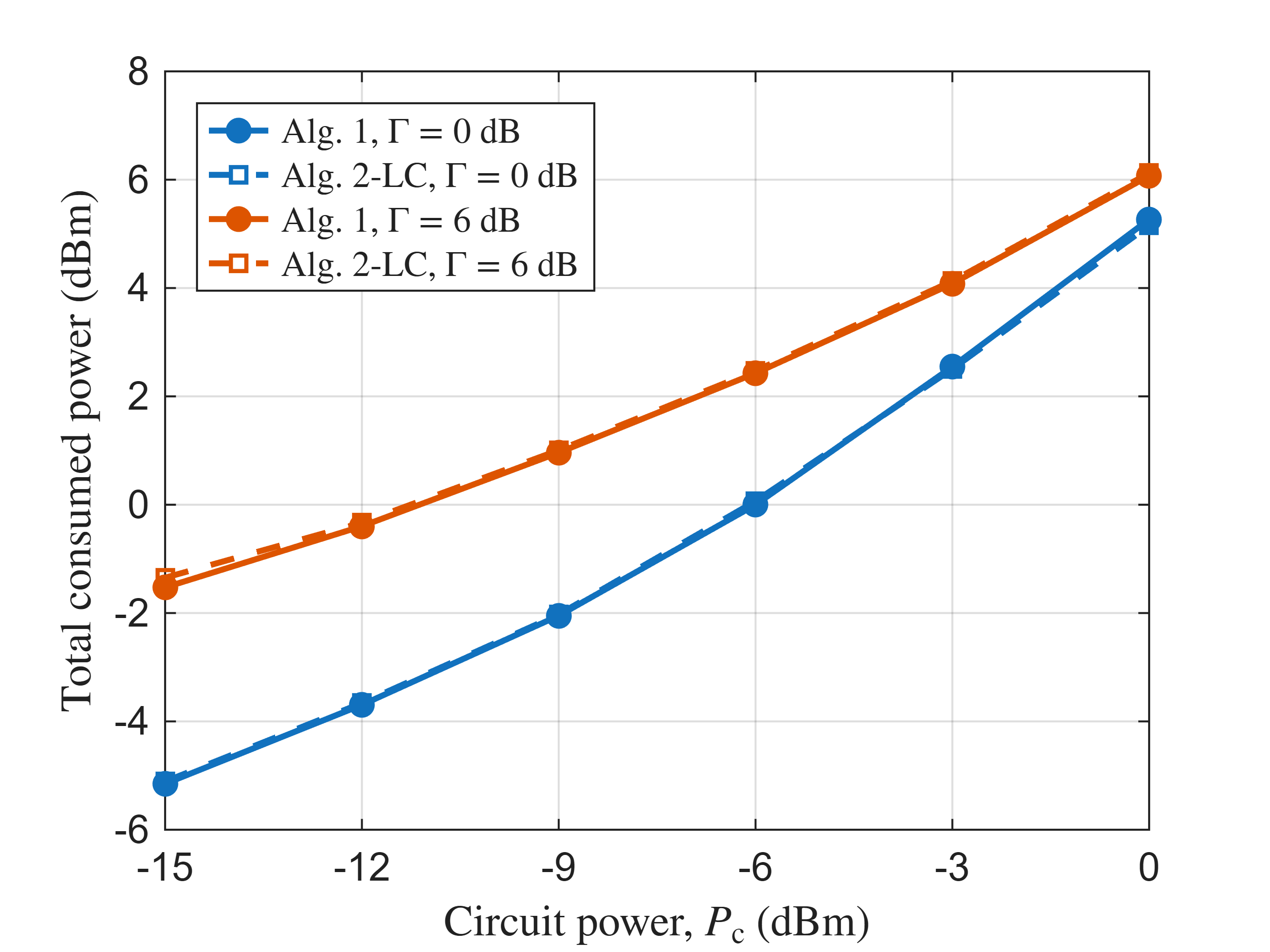}\\
        \captionsetup{justification=justified, singlelinecheck=false, font=small}	
        \caption{Total consumed power versus circuit power $P_{\rm c}$ for Algorithm~\ref{alg:dl_group_sparse} and Algorithm~\ref{alg:geometry_guided_dl} under different SINR requirements.} \label{fig:alg_compare_pc} 
\end{figure} 
Fig.~\ref{fig:alg_compare_pc} compares the total consumed power achieved by Algorithm~\ref{alg:dl_group_sparse} and Algorithm~\ref{alg:geometry_guided_dl} versus the circuit power $P_{\rm c}$ under $\Gamma=0$ dB and $\Gamma=6$ dB. For both algorithms, the consumed power increases with $P_{\rm c}$, since activating each APU becomes more costly. The curves for $\Gamma=6$ dB are higher than those for $\Gamma=0$ dB, as a stricter SINR requirement requires higher transmit power and/or more active APUs.
It can also be observed that Algorithm~\ref{alg:geometry_guided_dl} almost overlaps with Algorithm~\ref{alg:dl_group_sparse} for both SINR targets. This shows that the geometry-guided candidate-pool construction can identify the most useful APUs with negligible consumed-power loss. Therefore, Algorithm~\ref{alg:geometry_guided_dl} provides a low-complexity alternative to the full reweighted sparse beamforming design while preserving nearly the same energy-efficiency performance.

\begin{table}[t]
\centering
\caption{Average runtime comparison between Algorithm~\ref{alg:dl_group_sparse} and Algorithm~\ref{alg:geometry_guided_dl} under different numbers of APUs.}
\renewcommand{\arraystretch}{1.15}
\setlength{\tabcolsep}{4pt}
\label{tab:runtime_algorithms}
\begin{tabular}{
>{\centering\arraybackslash}p{0.1\linewidth}|
>{\centering\arraybackslash}p{0.26\linewidth}|
>{\centering\arraybackslash}p{0.26\linewidth}|
>{\centering\arraybackslash}p{0.2\linewidth}}
\hline
\hline
\rowcolor{blue!10}
\multicolumn{4}{c}{$\Gamma=0$ dB}\\
\hline
$N$  & Alg.~1 runtime (s) & Alg.~2 runtime (s) & Speedup \\
\hline
8 & 10.235 & 3.659 & 2.80$\times$ \\
12 & 15.399 & 3.592 & 4.29$\times$ \\
16 & 22.201 & 3.597 & 6.17$\times$ \\
20 & 29.396 & 3.605 & 8.15$\times$ \\
24 & 36.470 & 3.615 & 10.09$\times$ \\
\hline
\rowcolor{blue!12}
\multicolumn{4}{c}{$\Gamma=6$ dB}\\
\hline
$N$ & Alg.~1 runtime (s) & Alg.~2 runtime (s) & Speedup \\
\hline
8 & 9.699 & 3.646 & 2.66$\times$ \\
12 & 15.603 & 3.651 & 4.27$\times$ \\
16 & 21.402 & 3.526 & 6.07$\times$ \\
20 & 29.104 & 3.611 & 8.06$\times$ \\
24 & 41.329 & 4.995 & 8.27$\times$ \\
\hline
\hline
\end{tabular}
\end{table}
Table~\ref{tab:runtime_algorithms} compares the average runtime of Algorithm~\ref{alg:dl_group_sparse} and Algorithm~\ref{alg:geometry_guided_dl} under different numbers of deployed APUs. Two target SINR requirements, i.e., $\Gamma=0$ dB and $\Gamma=6$ dB, are considered. The speedup is defined as the ratio between the runtime of Algorithm~\ref{alg:dl_group_sparse} and that of Algorithm~\ref{alg:geometry_guided_dl}. It is observed that Algorithm~\ref{alg:geometry_guided_dl} consistently requires much less runtime, and the speedup becomes more significant as $N$ increases. For example, when $\Gamma=0$ dB, the speedup increases from $2.80\times$ at $N=8$ to $10.09\times$ at $N=24$. Similarly, when $\Gamma=6$ dB, the speedup increases from $2.66\times$ to $8.27\times$.
This runtime advantage comes from the reduced search space of Algorithm~\ref{alg:geometry_guided_dl}. Specifically, Algorithm~\ref{alg:dl_group_sparse} solves reweighted sparse beamforming problems over all deployed APUs, whose complexity grows with $N$. In contrast, Algorithm~\ref{alg:geometry_guided_dl} first constructs a fixed-size geometry-guided candidate pool and then performs beamforming and pruning within this smaller set. Therefore, its runtime grows much more slowly with $N$, making it more suitable for dense RS-GPA deployments.

\subsection{Performance Evaluation for Uplink Transmission}

\begin{figure}[!t]
	\centering
	\includegraphics[width=0.92\linewidth]{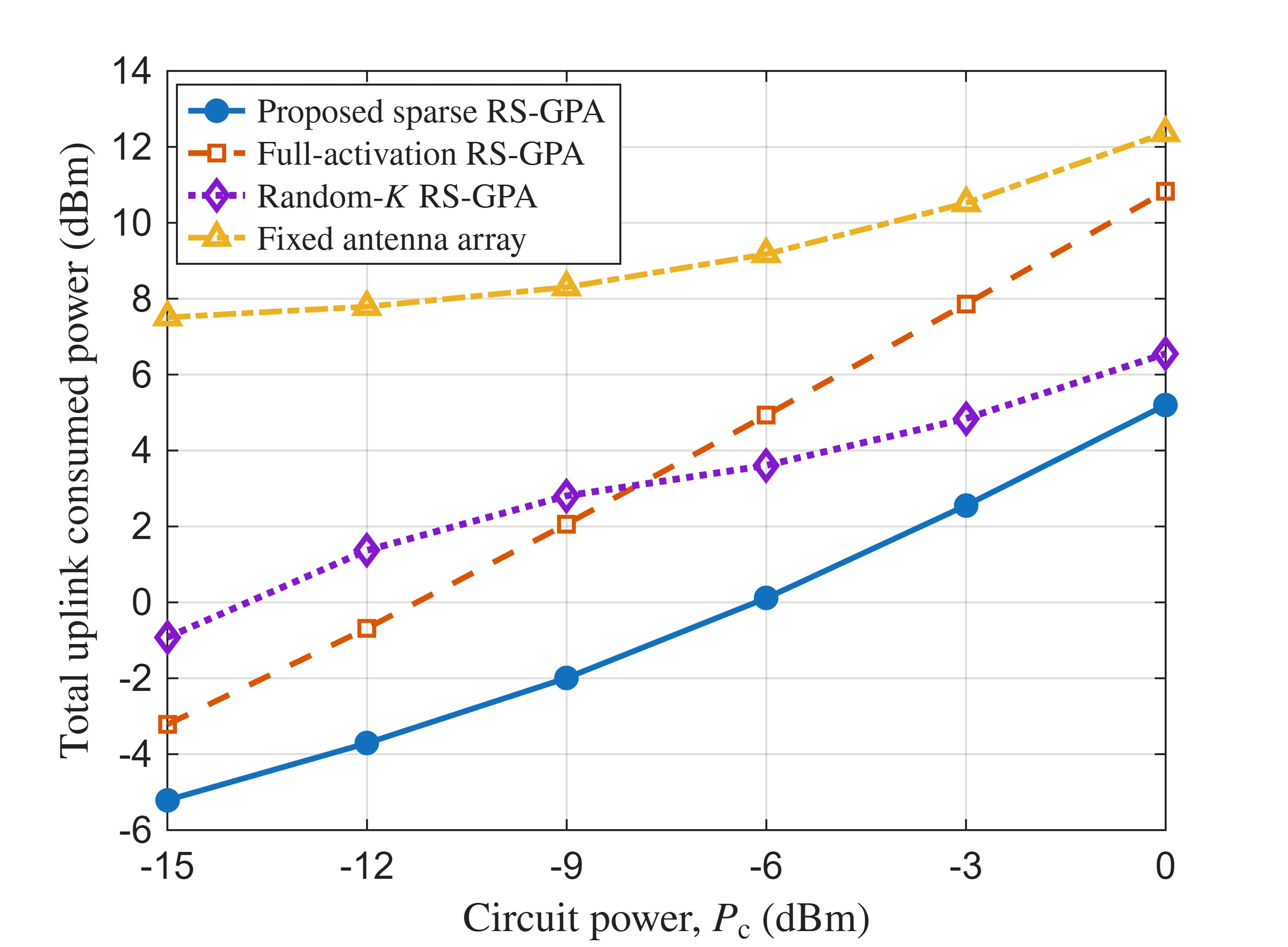}\\
        \captionsetup{justification=justified, singlelinecheck=false, font=small}	
        \caption{Total uplink consumed power versus circuit power $P_{\rm c}$ for different uplink transmission schemes.} \label{fig:ul_pc_power} 
\end{figure} 

Fig.~\ref{fig:ul_pc_power} shows the total uplink consumed power versus the circuit power $P_{\rm c}$. Four schemes are compared, where the random-$K$ RS-GPA scheme activates the same number of APUs as the proposed sparse RS-GPA scheme but selects them randomly. It can be observed that the consumed power of all schemes generally increases with $P_{\rm c}$, since each activated receiving unit incurs a larger circuit-power cost.
The proposed sparse RS-GPA scheme achieves the lowest consumed power over the whole considered range. This demonstrates that the proposed uplink design can effectively balance the reduction in user transmit power and the circuit-power cost of APU activation. The full-activation RS-GPA scheme performs well when $P_{\rm c}$ is small, but its consumed power increases rapidly as $P_{\rm c}$ grows because all APUs remain active. The random-$K$ RS-GPA scheme also consumes more power than the proposed scheme, confirming the importance of optimized APU selection. The fixed antenna array has the highest consumed power in most cases, since its fixed antenna locations cannot exploit the spatial flexibility of the RS.

\begin{figure}[!t]
	\centering
	\includegraphics[width=0.92\linewidth]{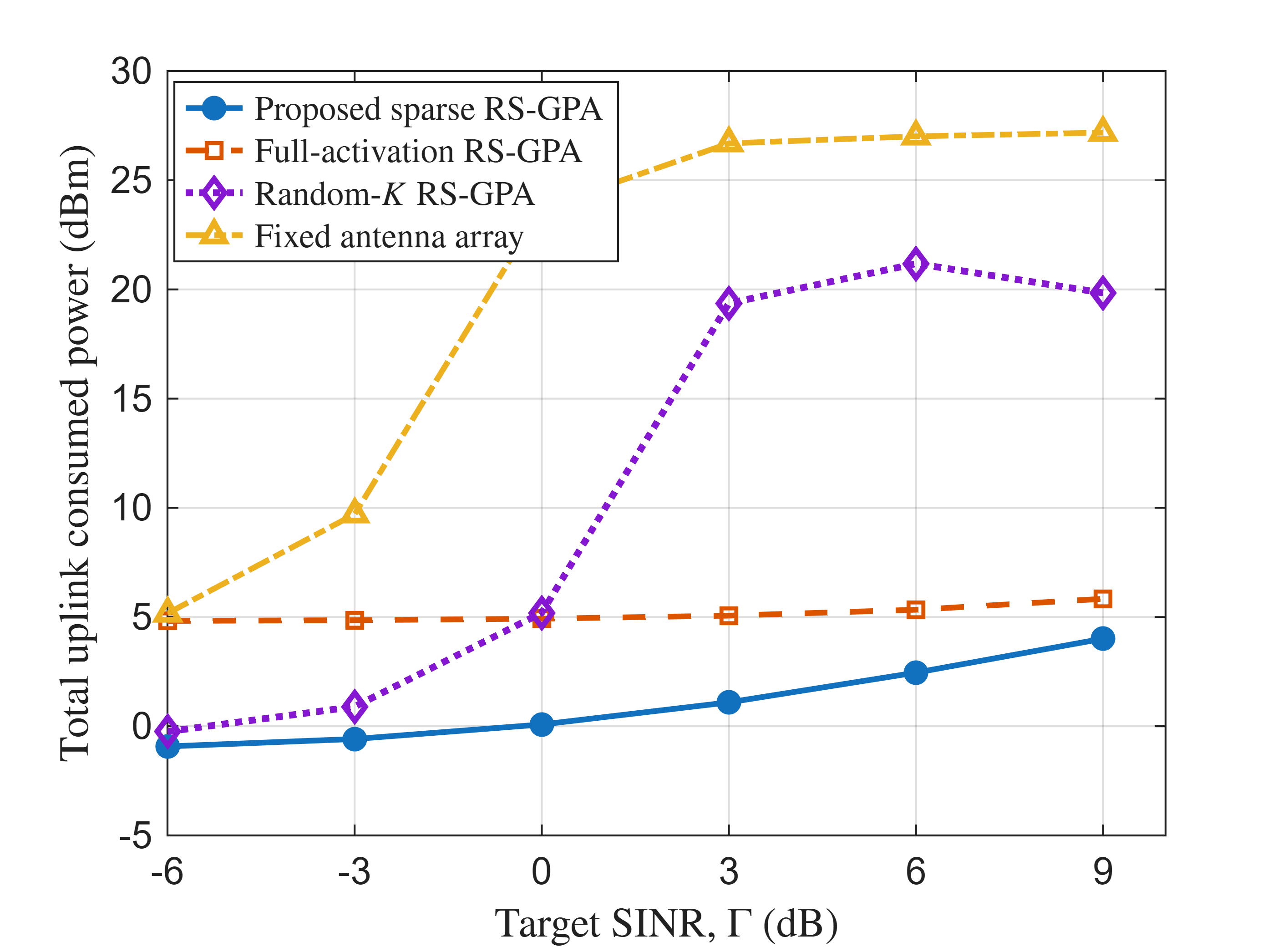}\\
        \captionsetup{justification=justified, singlelinecheck=false, font=small}	
        \caption{Total uplink consumed power versus target SINR $\Gamma$ for different uplink transmission schemes.} \label{fig:ul_power_vs_sinr} 
\end{figure} 

Fig.~\ref{fig:ul_power_vs_sinr} shows the total uplink consumed power versus the target SINR $\Gamma$. The proposed sparse RS-GPA scheme achieves the lowest consumed power over the whole SINR range, demonstrating that the proposed uplink design can effectively balance user transmit power and active-APU circuit power. As $\Gamma$ increases, the consumed power of the proposed scheme increases moderately, since a higher SINR requirement requires stronger received signal quality.
The full-activation RS-GPA scheme has higher consumed power at low SINR due to the circuit-power cost of activating all APUs, while the gap to the proposed scheme becomes smaller at high SINR because more APUs are useful for reducing user transmit powers. The random-$K$ RS-GPA scheme performs much worse in the medium-to-high SINR regime, showing that randomly selected APUs may fail to provide effective receive gain and interference suppression. The fixed antenna array requires the highest consumed power in most cases, especially for moderate and large SINR requirements, due to its lack of spatial flexibility. These results verify the effectiveness of sparse APU activation and uplink power control for energy-efficient RS-GPA transmission.

\section{Conclusion} \label{sec: conclusion}

In this paper, we proposed an RS-GPA framework, where the APUs deployed along an RS are interpreted as discrete and controllable radiation or reception points for location-flexible wireless access. Based on this framework, we established the downlink and uplink signal models and investigated circuit-power-aware sparse APU activation for energy-efficient communications. For downlink transmission, we formulated a total consumed power minimization problem and developed a reweighted group-sparse beamforming algorithm. We further analyzed the single-user case to reveal the APU activation principle and developed a geometry-guided low-complexity multiuser algorithm. For uplink transmission, we formulated a joint APU activation and user power control problem and proposed an efficient sparse activation design. Numerical results demonstrated that the proposed RS-GPA designs can substantially reduce the total consumed power compared with benchmark schemes, while the geometry-guided algorithm achieves near-identical performance to the group-sparse design with significantly lower computational complexity. These results confirm the potential of RS-GPA systems as a practical platform for energy-efficient and location-flexible wireless access.


\smaller[1]


\begin{thebibliography}{10}
	\providecommand{\url}[1]{#1}
	\csname url@samestyle\endcsname
	\providecommand{\newblock}{\relax}
	\providecommand{\bibinfo}[2]{#2}
	\providecommand{\BIBentrySTDinterwordspacing}{\spaceskip=0pt\relax}
	\providecommand{\BIBentryALTinterwordstretchfactor}{4}
	\providecommand{\BIBentryALTinterwordspacing}{\spaceskip=\fontdimen2\font plus
		\BIBentryALTinterwordstretchfactor\fontdimen3\font minus \fontdimen4\font\relax}
	\providecommand{\BIBforeignlanguage}[2]{{%
			\expandafter\ifx\csname l@#1\endcsname\relax
			\typeout{** WARNING: IEEEtran.bst: No hyphenation pattern has been}%
			\typeout{** loaded for the language `#1'. Using the pattern for}%
			\typeout{** the default language instead.}%
			\else
			\language=\csname l@#1\endcsname
			\fi
			#2}}
	\providecommand{\BIBdecl}{\relax}
	\BIBdecl
	
	\bibitem{you2021towards}
	X.~You, C.-X. Wang, J.~Huang, X.~Gao, Z.~Zhang, M.~Wang, Y.~Huang, C.~Zhang, Y.~Jiang, J.~Wang \emph{et~al.}, ``Towards {6G} wireless communication networks: Vision, enabling technologies, and new paradigm shifts,'' \emph{Sci. China Inf. Sci.}, vol.~64, no.~1, p. 110301, 2021.
	
	\bibitem{xu2025distributed}
	Y.~Xu, E.~G. Larsson, E.~A. Jorswieck, X.~Li, S.~Jin, and T.-H. Chang, ``Distributed signal processing for extremely large-scale antenna array systems: State-of-the-art and future directions,'' \emph{IEEE J. Sel. Top. Signal Process.}, vol.~19, no.~2, pp. 304--330, 2025.
	
	\bibitem{wu2019intelligent}
	Q.~Wu and R.~Zhang, ``Intelligent reflecting surface enhanced wireless network via joint active and passive beamforming,'' \emph{IEEE Trans. Wireless Commun.}, vol.~18, no.~11, pp. 5394--5409, 2019.
	
	\bibitem{shao20246d}
	X.~Shao, Q.~Jiang, and R.~Zhang, ``{6D} movable antenna based on user distribution: Modeling and optimization,'' \emph{IEEE Trans. Wireless Commun.}, vol.~24, no.~1, pp. 355--370, 2024.
	
	\bibitem{shao20256d}
	X.~Shao, R.~Zhang, Q.~Jiang, and R.~Schober, ``{6D} movable antenna enhanced wireless network via discrete position and rotation optimization,'' \emph{IEEE J. Sel. Areas Commun.}, vol.~43, no.~3, pp. 674--687, 2025.
	
	\bibitem{zhu2023modeling}
	L.~Zhu, W.~Ma, and R.~Zhang, ``Modeling and performance analysis for movable antenna enabled wireless communications,'' \emph{IEEE Trans. Wireless Commun.}, vol.~23, no.~6, pp. 6234--6250, 2023.
	
	\bibitem{wong2020fluid}
	K.-K. Wong, A.~Shojaeifard, K.-F. Tong, and Y.~Zhang, ``Fluid antenna systems,'' \emph{IEEE Trans. Wireless Commun.}, vol.~20, no.~3, pp. 1950--1962, 2020.
	
	\bibitem{suzuki2022pinching}
	A.~Fukuda, H.~Yamamoto, H.~Okazaki, Y.~Suzuki, and K.~Kawai, ``Pinching antenna: Using a dielectric waveguide as an antenna,'' \emph{NTT DOCOMO Technical J.}, vol.~23, no.~3, pp. 5--12, Jan. 2022.
	
	\bibitem{ding2025flexible}
	Z.~Ding, R.~Schober, and H.~V. Poor, ``Flexible-antenna systems: A pinching-antenna perspective,'' \emph{IEEE Trans. Commun.}, vol.~73, no.~10, pp. 9236--9253, Oct. 2025.
	
	\bibitem{xu2025rate}
	Y.~Xu, Z.~Ding, and G.~K. Karagiannidis, ``Rate maximization for downlink pinching-antenna systems,'' \emph{IEEE Wireless Commun. Lett.}, vol.~14, no.~5, pp. 1431--1435, May 2025.
	
	\bibitem{qian2026pinching}
	M.~Qian, X.~Mu, L.~You, and M.~Matthaiou, ``Pinching-antenna-based communications: Spectral efficiency analysis and deployment strategies,'' \emph{IEEE Trans. Commun.}, vol.~74, pp. 4758--4771, 2026.
	
	\bibitem{li2026power}
	L.~Li, Y.~Xu, T.~Cai, and T.-H. Chang, ``Power minimization in pinching-antenna systems under probabilistic {LoS} blockage,'' \emph{IEEE Wireless Commun. Lett.}, vol.~15, pp. 3139--3143, 2026.
	
	\bibitem{zhou2026spectral}
	Z.~Zhou, Z.~Wang, and Y.~Liu, ``Spectral and energy efficiency tradeoff for pinching-antenna systems,'' \emph{IEEE Trans. Commun.}, vol.~74, pp. 8283--8299, 2026.
	
	\bibitem{xu2026environment1}
	Y.~Xu, Z.~Ding, X.~Y. Zhang, T.~Q. Duong, and T.-H. Chang, ``Environment-aware network-level design of generalized pinching-antenna systems--part {I}: Traffic-aware case,'' \emph{arXiv preprint arXiv:2602.17023}, 2026.
	
	\bibitem{xu2026environment2}
	------, ``Environment-aware network-level design of generalized pinching-antenna systems--part {II}: Geometry-aware case,'' \emph{arXiv preprint arXiv:2602.17032}, 2026.
	
	\bibitem{tyrovolas2025performance}
	D.~Tyrovolas, S.~A. Tegos, P.~D. Diamantoulakis, S.~Ioannidis, C.~K. Liaskos, and G.~K. Karagiannidis, ``Performance analysis of pinching-antenna systems,'' \emph{IEEE Trans. Cognit. Commun. Networking}, vol.~12, pp. 590--601, 2026.
	
	\bibitem{mao2025multi}
	W.~Mao, Y.~Lu, Y.~Xu, B.~Ai, O.~A. Dobre, and D.~Niyato, ``Multi-waveguide pinching antennas for {ISAC},'' \emph{IEEE Trans. Wireless Commun.}, vol.~25, pp. 5846--5858, 2026.
	
	\bibitem{ouyang2026rate}
	C.~Ouyang, Z.~Wang, Y.~Liu, and Z.~Ding, ``Rate region of {ISAC} for pinching-antenna systems,'' \emph{IEEE Trans. Commun.}, vol.~74, pp. 5849--5866, 2026.
	
	\bibitem{xu2025qos}
	Y.~Xu, Z.~Ding, D.~Cai, and V.~W. Wong, ``{QoS}-aware {NOMA} design for downlink pinching-antenna systems,'' \emph{IEEE Trans. Commun.}, vol.~73, no.~12, pp. 13\,611--13\,625, 2025.
	
	\bibitem{gan2025joint}
	D.~Gan, X.~Xu, J.~Zuo, X.~Ge, and Y.~Liu, ``Joint beamforming for {NOMA} assisted pinching antenna systems ({PASS}),'' \emph{IEEE Trans. Commun.}, vol.~74, pp. 2450--2465, 2025.
	
	\bibitem{papanikolaou2025resolving}
	V.~K. Papanikolaou, G.~Zhou, B.~Kaziu, A.~Khalili, P.~D. Diamantoulakis, G.~K. Karagiannidis, and R.~Schober, ``Resolving the double near-far problem via wireless powered pinching-antenna networks,'' \emph{IEEE Wireless Commun. Lett.}, vol.~14, no.~11, pp. 3425--3429, 2025.
	
	\bibitem{hu2026average}
	H.~Hu, R.~Jiang, Y.~Xu, J.~Ma, and F.~Fang, ``Average aoi in pinching antenna-assisted wpcns with probabilistic los blockage,'' \emph{IEEE Commun. Lett.}, vol.~30, pp. 1786--1790, 2026.
	
	\bibitem{zhong2025physical}
	Y.~Zhong, J.~Chen, Y.~Xiao, S.~Yang, X.~Lei, Y.~Gao, and M.~Xiao, ``Physical layer security for pinching-antenna systems via index and directional modulation,'' \emph{IEEE Wireless Commun. Lett.}, vol.~15, pp. 230--234, 2026.
	
	\bibitem{zhu2025pinching}
	G.~Zhu, X.~Mu, L.~Guo, S.~Xu, Y.~Liu, and N.~Al-Dhahir, ``Pinching-antenna systems ({PASS})-enabled secure wireless communications,'' \emph{IEEE Trans. Commun.}, vol.~74, pp. 490--505, 2026.
	
	\bibitem{xu2025generalized}
	Y.~Xu, J.~Cui, Y.~Zhu, Z.~Ding, T.-H. Chang, R.~Schober, V.~W. Wong, O.~A. Dobre, G.~K. Karagiannidis, H.~V. Poor, and X.~You, ``Generalized pinching-antenna systems: A tutorial on principles, design strategies, and future directions,'' \emph{IEEE Commun. Surv. Tutorials}, vol.~28, pp. 5872--5908, 2026.
	
	\bibitem{wang2025generalized}
	K.~Wang, Z.~Ding, and L.~Hanzo, ``Generalized pinching-antenna systems: A leaky-coaxial-cable perspective,'' \emph{arXiv preprint arXiv:2512.04979}, 2025.
	
	\bibitem{wang2026leaky1}
	K.~Wang, Z.~Ding, and D.~K. So, ``Leaky coaxial cable based generalized pinching-antenna systems with dual-port feeding,'' \emph{arXiv preprint arXiv:2602.21856}, 2026.
	
	\bibitem{wang2026leaky2}
	K.~Wang, D.~K. So, Z.~Ding, and G.~K. Karagiannidis, ``Leaky-coaxial pinching-antenna system with adjustable slot apertures,'' \emph{arXiv preprint arXiv:2604.23246}, 2026.
	
	\bibitem{interdonato2019ubiquitous}
	G.~Interdonato, E.~Bj{\"o}rnson, H.~Quoc~Ngo, P.~Frenger, and E.~G. Larsson, ``Ubiquitous cell-free massive {MIMO} communications,'' \emph{EURASIP J. Wireless Commun. Networking}, vol. 2019, no.~1, pp. 1--13, 2019.
	
	\bibitem{shaik2021mmse}
	Z.~H. Shaik, E.~Bj{\"o}rnson, and E.~G. Larsson, ``{MMSE}-optimal sequential processing for cell-free massive {MIMO} with radio stripes,'' \emph{IEEE Trans. Commun.}, vol.~69, no.~11, pp. 7775--7789, 2021.
	
	\bibitem{lopez2022massive}
	O.~L. L{\'o}pez, D.~Kumar, R.~D. Souza, P.~Popovski, A.~T{\"o}lli, and M.~Latva-Aho, ``Massive {MIMO} with radio stripes for indoor wireless energy transfer,'' \emph{IEEE Trans. Wireless Commun.}, vol.~21, no.~9, pp. 7088--7104, 2022.
	
	\bibitem{chiotis2024uplink}
	I.~Chiotis and A.~L. Moustakas, ``Uplink performance optimization of limited-capacity radio stripes,'' \emph{IEEE Trans. Wireless Commun.}, vol.~23, no.~9, pp. 12\,382--12\,395, 2024.
	
	\bibitem{conceiccao2023access}
	F.~Concei{\c{c}}{\~a}o, L.~Martins, M.~Gomes, V.~Silva, and R.~Dinis, ``Access point selection for spectral efficiency and load balancing optimization in radio stripes,'' \emph{IEEE Commun. Lett.}, vol.~27, no.~9, pp. 2383--2387, 2023.
	
	\bibitem{eberechukwu2025radio}
	P.~Eberechukwu and D.~Yoon, ``Radio stripe selection for improved {UE} position estimation in {CF-mMIMO} systems,'' \emph{IEEE Trans. Veh. Technol.}, pp. 1--6, 2025.
	
	\bibitem{xu2025pinching}
	Y.~Xu, Z.~Ding, R.~Schober, and T.-H. Chang, ``Pinching-antenna systems with in-waveguide attenuation: Performance analysis and algorithm design,'' \emph{IEEE Trans. Wireless Commun.}, vol.~25, pp. 14\,564--14\,580, 2026.
	
	\bibitem{gao2017massive}
	Y.~Gao, H.~Vinck, and T.~Kaiser, ``Massive {MIMO} antenna selection: Switching architectures, capacity bounds, and optimal antenna selection algorithms,'' \emph{IEEE Trans. Signal Process.}, vol.~66, no.~5, pp. 1346--1360, 2017.
	
	\bibitem{asaad2018massive}
	S.~Asaad, A.~M. Rabiei, and R.~R. M{\"u}ller, ``Massive {MIMO} with antenna selection: Fundamental limits and applications,'' \emph{IEEE Trans. Wireless Commun.}, vol.~17, no.~12, pp. 8502--8516, 2018.
	
	\bibitem{dai2014sparse}
	B.~Dai and W.~Yu, ``Sparse beamforming and user-centric clustering for downlink cloud radio access network,'' \emph{IEEE Access}, vol.~2, pp. 1326--1339, 2014.
	
	\bibitem{candes2008enhancing}
	E.~J. Candes, M.~B. Wakin, and S.~P. Boyd, ``Enhancing sparsity by reweighted $\ell_1$ minimization,'' \emph{J.Fourier Anal. Appl.}, vol.~14, no.~5, pp. 877--905, 2008.
	
\end{thebibliography}
\end{document}